\documentclass[11pt]{amsart}
\usepackage{amsbsy,amssymb,amscd,amsfonts,latexsym,amstext,delarray,
amsmath,graphicx} 
\input xypic

\usepackage[utf8]{inputenc}
\usepackage{multicol}
\usepackage{subfig}
\usepackage{float}
\usepackage{dblfloatfix}
\usepackage{subfig}
\usepackage{url}
\usepackage{dsfont}
\usepackage{indentfirst}
\usepackage{caption}

\usepackage{tikz-cd}

\usepackage{color}

\newtheorem{thm}{Theorem}[section]
\newtheorem{prop}[thm]{Proposition}
\newtheorem{cor}[thm]{Corollary}
\newtheorem{lem}[thm]{Lemma}

\newtheorem{defn}[thm]{Definition}

\newtheorem{ex}[thm]{Example}

\numberwithin{equation}{section}

\def\bS{{\mathbb S}}

\def\C{{\mathbb C}}

\renewcommand{\H}{{\mathbb H}}
\def\N{{\mathbb N}}
\renewcommand{\P}{{\mathbb P}}

\def\R{{\mathbb R}}
\def\Z{{\mathbb Z}}

\def\cancel#1#2{\ooalign{$\hfil#1\mkern1mu/\hfil$\crcr$#1#2$}}
\def\Dirac{\mathpalette\cancel D}

\def\cA{{\mathcal A}}

\def\cC{{\mathcal C}}
\def\cD{{\mathcal D}}

\def\cG{{\mathcal G}}
\def\cH{{\mathcal H}}
\def\cI{{\mathcal I}}

\def\cK{{\mathcal K}}
\def\cL{{\mathcal L}}

\def\cS{{\mathcal S}}

\def\GL{{\rm GL}}

\def\PSL{{\rm PSL}}

\def\SL{{\rm SL}}
\def\Spec{{\rm Spec}}

\def\Tr{{\rm Tr}}

\def\fs{{\mathfrak s}}

\DeclareMathOperator{\arccosh}{arccosh}

\title{Adinkras, Dessins, Origami, and Supersymmetry Spectral Triples}
\author{Matilde Marcolli and Nick Zolman}
\address{Division of Physics, Mathematics, and Astronomy, California Institute of Technology, 1200 E California Blvd, Pasadena, CA 91125, USA}
\email{matilde@caltech.edu}
\email{nzolman@caltech.edu}

\begin{document}
\maketitle

\begin{abstract}
We investigate the spectral geometry and spectral action functionals associated
to 1D Supersymmetry Algebras, using the classification of these superalgebras
in terms of Adinkra graphs and the construction of associated dessin d'enfant and
origami curves.  The resulting spectral action functionals are computed in terms
of the Selberg (super) trace formula. 
\end{abstract}

\section{Introduction}

In this paper we construct spectral geometries (spectral triples) and 
spectral action functionals associated to 1D supersymmetry algebras.
We use the classification of \cite{FaGa} of these superalgebras in
terms of combinatorial data of Adinkra graphs, in combination with
recent results of \cite{DILM}, which show that an Adinkra graph
determines geometric data:
\begin{itemize}
\item a compact Riemann surface $X$, 
\item a branched covering $f: X \to \P^1(\C)$ ramified at $\{ 0,1,\infty \}$ (that is, a Belyi map), 
\item a spin structure on $X$,
\item a  structure of Super Riemann Surface on $X$.
\end{itemize}
The first two property specify a {\em dessin d'enfant}, in the sense of Grothendieck, 
given by the graph $f^{-1}(\cI)$, for $\cI=[0,1]$, embedded in $X$.

\smallskip

We associate a spectral geometry to a
given 1D supersymmetry algebra, by considering the canonical
spectral triple $(\cC^\infty(X), L^2(X,\bS_\fs, \Dirac_\fs)$ on the
Riemann surface $X$, with spin structure $\fs$ and spinor 
bundle $\bS_\fs$ and Dirac operator $\Dirac_\fs$, where $X$ and
$\fs$ are determined by the Adinkra of the 1D supersymmetry algebra,
as in \cite{DILM}. 

\smallskip

We show that the associate spectral action functional
of the spectral triple can be computed using the Selberg trace formula
for a finite index subgroup of a Fuchsian triangle group of the form
$\Delta_{N,N,2}$. For a special choice of the test function, the
spectral action functional can be computed in terms of the Selberg
zeta function, which in turn can be computed with the method of
\cite{Pohl}, using the thermodynamic formalism and the Ruelle transfer
operator of a symbolic dynamics coding the geodesic flow on $X$. 

\smallskip

We also show that, given the transfer operator for the Fuchsian triangle group
$\Delta_{N,N,2}$, the one for a finite index subgroup $H\subset \Delta_{N,N,2}$ 
can be obtained by extending the symbolic dynamics and the transfer operator 
from the boundary of the upper half plane to its product with the coset space
$P=\Delta_{N,N,2}/H$. 

\smallskip

The results of \cite{DILM} show that the Adinkra classifying the 
1D supersymmetry algebra also determines on $X$ the structure of a 
Super Riemann Surface. We refine the construction of the spectral
geometry and the spectral action functional by incorporating this
supersymmetry, and replacing the Dirac operator with the supersymmetric
Dirac Laplacian, which is an operator proportional to $D \bar D$, where
$D=\partial_\theta + \theta \partial_z$ on the Super Riemann Surface. 
The associated action functional is a super spectral action computed
as a supertrace of this operator, regularized by a test function. We show
that this action functional can be computed using the Selberg
supertrace formula of \cite{BarMan}. 

\smallskip

We also show, by using the construction in \cite{Moll} of origami curves associated
to dessins d'enfant, that the Adinkra graph $A$ of a 1D supersymmetry algebra
determines uniquely an origami curve $Y$, in which the graph $A$ embeds,
with a choice of $2^n$ embeddings, where $n=\# E(A)$ is the number of edges. 
We then use a result of \cite{Mato2} showing that, for all Riemann surfaces
$Y$ that admits a branched cover $p: Y \to E$, with $E$ an elliptic curve
(hence in particular for all origami curves), it is possible to construct a family
of metrics on $Y$, determined by compatibility conditions on the period matrix
coming from the existence of the branched cover map to $E$. For each
metric in this family, it is shown in \cite{Mato2} how to obtain an infinite family of eigenvalues 
and eigenvectors of the corresponding Laplacian. The resulting spectrum and spectral action
functional behave more like the spectral action of tori and its computation
can be approached in terms of a Poisson summation formula.

\smallskip
\subsection{Supersymmetry and Spectral geometry}

Noncommutative geometry has developed a broad approach to the
construction of particle physics models and gravity models (see the
overviews given in \cite{WvS} and in the upcoming \cite{Mar2}). These
model provide extensions of the minimal standard model of elementary
particle physics, as well as modified gravity models in cosmology,
obtained from the spectral action functional of a suitable (noncommutative) geometry.
Among the extensions of the minimal standard model that have been
considered so far with this method, the recent work \cite{BBS} showed
to what extent the Minimally Supersymmetric Standard Model (MSSM) can be
incorporated and recovered from this general approach. The results
of \cite{BBS} provided a novel and important insight on noncommutative
geometry and supersymmetry. 

\smallskip

In the present paper, we consider a different type of supersymmetric
models, much simpler in nature than the MSSM of particle physics,
namely the 1-dimensional supersymmetry algebras, or $(1|N)$-Superalgebras.

\smallskip

We use the classification of \cite{FaGa} in terms of Adinkra graphs, and
the results of \cite{DILM} associating to an Adinkra graph a dessin d'enfant
and a Belyi map, in order to associate to a supersymmetry algebra a 
geometric object, a Riemann surface, or a Super Riemann Surface with a
spin structure. We then use this geometric object to obtain a spectral triple
associated to the supersymmetry algebra. It is given by the standard spectral
triple of this spin geometry. We compute the associated spectral action
functional. We also consider another construction, which is based on
the relation between dessins d'enfant and origami curves as in \cite{Moll},
to construct another related geometry and compare their spectral properties. 

\smallskip

While the models we consider here are not directly related to the
particle physics models considered in \cite{BBS}, they provide a rich
class of examples of physical models with supersymmetry that have
an associated spectral geometry and spectral action functional. It would 
be interesting to further investigate how this point of view based on
spectral triples and the spectral action relates, for example, to the
real homotopy theory approach of \cite{KimSa}, though this
is beyond the scope of the present paper.

\smallskip
\subsection{Supersymmetry Algebras}

In the theory of Supersymmetry, the possible symmetries of a $4$-dimensional quantum field theory
(viewed as symmetries of the $S$-matrix) are classified by the Haag--Lopuszanski--Sohnius theorem, \cite{HaSoLo}.
This result shows that the possible symmetries consist of internal symmetries and a nontrivial extension
of the Poincar\'e algebra: the supersymmetry algebras. 

\smallskip

The off-shell supersymmetry algebras we will be considering here correspond to the
case of a $1$-dimensional space-time, with time coordinate $t$, and zero-dimensional space
(supersymmetric quantum mechanics). We consider the $(1|N)$ Superalgebras, generated by 
operators  $Q_1, Q_2, \dots, Q_N$ and $H$, where the $Q_k$, $k=1,\ldots, N$ are the 
supersymmetry generators, and $H=i\partial_t$.  These operators are required to satisfy the following
commutation relations:
			\begin{gather}
				[Q_k, H] = 0 \label{commutation} \\ 
				\{Q_k, Q_\ell\} = 2 \delta_{k\ell} H. \label{anticommutation}
			\end{gather}	
			
We consider representations of these operators, acting on bosonic and fermionic fields. 
Let $\{\phi_1, \ldots, \phi_m \}$ be a set of bosonic fields, given by commuting real valued functions, 
and $\{\psi_1, \dots, \psi_m \}$ be a set of fermionic fields, given by anticommuting real 
(Grassman variable) valued functions. The off-shell condition means that the only relations
are given by \eqref{commutation} and \eqref{anticommutation} and no other equation holds imposing
additional relations between these fields. We consider operators acting as
\begin{align}
	Q_k \phi_a &= c \partial_{t}^\lambda \psi_b  
	\label{Q-action1} \\
	Q_k \psi_b &= \frac{i}{c} \partial_{t}^{1 -\lambda} \phi_a ,
	\label{Q-action2}
	\end{align}
for parameters $c \in \{-1,1\}$ and $\lambda \in \{0,1\}$. These representations clearly
satisfy the relations \eqref{commutation} and \eqref{anticommutation}.

\smallskip
\subsection{Adinkras}

Recently, a graphical method for classifying the
supersymmetry algebras was introduced by Faux and Gates, \cite{FaGa}. The resulting decorated bipartite
graphs are called Adinkras. The mathematical and physical properties of Adinkras were extensively studied
in recent years, see for instance \cite{DFGHIL} and \cite{DILM}. A good introduction to Adinkras for
mathematicians is given in \cite{Zhang}. We recall here briefly a few basic facts about Adinkras, 
following the aforementioned references.

\smallskip

Let $A$ be a finite graph that is simple, namely it has no looping edges and no parallel edges.
We denote by $V(A)$ and $E(A)$ the sets of vertices and edges of $A$.

\smallskip

An \textit{$N$-dimensional chromotopology} is a finite connected simple graph, $A$ such that:
			\begin{itemize}
				\item $A$ is $N$-regular (all vertices have valence $N$) and bipartite. 
				\item The elements of $E(A)$ are colored by $N$ colors, represented 
				by elements of the set $\{1, 2, \dots, N\}$.
				\item Every vertex is incident to exactly one edge of each color.
				\item For any distinct colors $i$ and $j$, the edges in $E(A)$ with 
				colors $i$ and $j$ form a disjoint union of 4-cycles.
			\end{itemize}
			
In superalgebras, the two sets of vertices that form the bipartition correspond to ``bosons" 
and ``fermions". We color them white and black, respectively. 

\smallskip

A \textit{ranking} on a graph $A$ is a partial ordering on the set $V(A)$, determined by
a function, $h: V(A) \rightarrow \mathbb{Z}$. One can represent the ranking on a graph 
as vertical placement of the vertices, that is, as a height function. 

\smallskip

A \textit{dashing} on a graph $A$ is a function $d: E(A) \rightarrow \Z/2\Z$ that assigns 
to each edge a value $0$ (solid) or $1$ (dashed). A $4$-cycle in a graph has an
\textit{odd-dashing} if it has an odd number of dashed edges. 
A colored graph whose $2$-colored $4$-cycles all have an odd-dashing is 
called \textit{well-dashed}.

\smallskip

An \textit{Adinkra} is a well-dashed, $N$-dimensional chromotopology, with a ranking 
on its bipartition, such that the bosons have even ranking and the fermions have odd ranking. 
We call any chromotopology \textit{Adinkraizable} if it can be well-dashed and it admits 
a well-defined ranking as above.  

\smallskip

The results of \cite{FaGa} showed that Adinkras can be used to classify the one-dimensional 
superalgebras and to generate large classes of significant examples of such algebras.  
We just recall here briefly how one encodes a superalgebra of the form \eqref{Q-action1} 
and \eqref{Q-action2} as an Adinkra graph. 

\smallskip

Given $\{\phi_1, \ldots, \phi_m \}$ (bosonic fields) and $\{\psi_1, \dots, \psi_m \}$ (fermionic fields)
with a representation of the form \eqref{Q-action1} and \eqref{Q-action2},
with $c \in \{-1,1\}$ and $\lambda \in \{0,1\}$, as above, one constructs an Adinkra graph with 
a set of white vertices corresponding to the set of bosonic fields and time derivatives and a set 
of black vertices corresponding to the fermionic fields and time derivatives. An edge is assigned
between a pair $\{ v_a , v_b \}$ of vertices 
for each relation of the form \eqref{Q-action1} or \eqref{Q-action2}, with the rule that the edge
is oriented from the white to the black vertex in the pair if $\lambda=0$ and from the 
black to the white vertex if $\lambda=1$. Moreover, the edge is dashed if $c=-1$ 
and solid if $c=1$. This orientation is obtained by effect of the ranking function $h: V(A)\to \Z$
described above. We can summarize these rules on ranking and dashing in the following table. 
\smallskip

\begin{center}
\begin{tabular}{ c c|c c } 

Action of $Q_I$ & Adinkra & Action of $Q_I$ & Adinkra \\  
\hline 

$Q_I \begin{bmatrix} \psi_B\\ \phi_A\end{bmatrix} = 
\begin{bmatrix} i \dot{\phi}_A\\ \psi_B\end{bmatrix}
$

& \raisebox{-0.5\totalheight}{\includegraphics[width=0.05\textwidth]{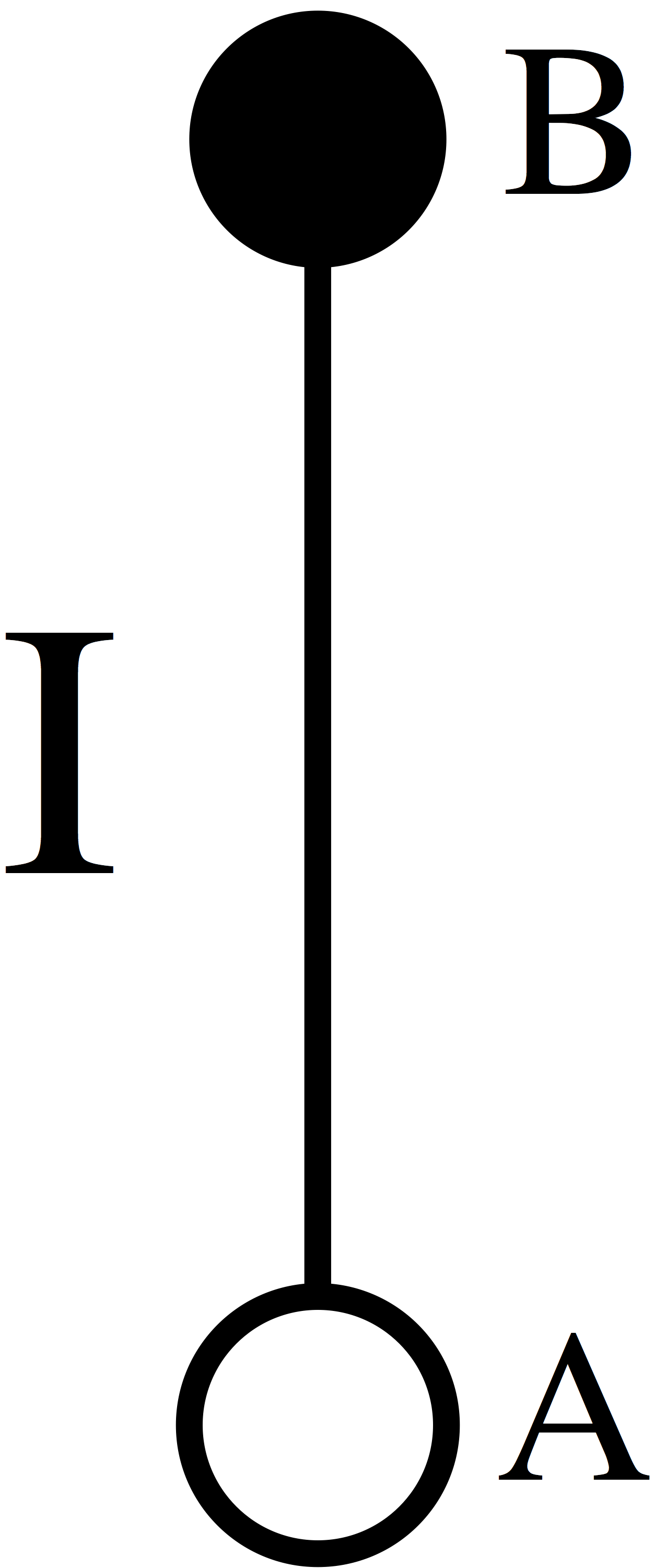}}

& $Q_I \begin{bmatrix} \psi_B\\ \phi_A\end{bmatrix} = 
\begin{bmatrix} -i \dot{\phi}_A\\ -\psi_B\end{bmatrix}
$

& \raisebox{-0.5\totalheight}{\includegraphics[width=0.05\textwidth]{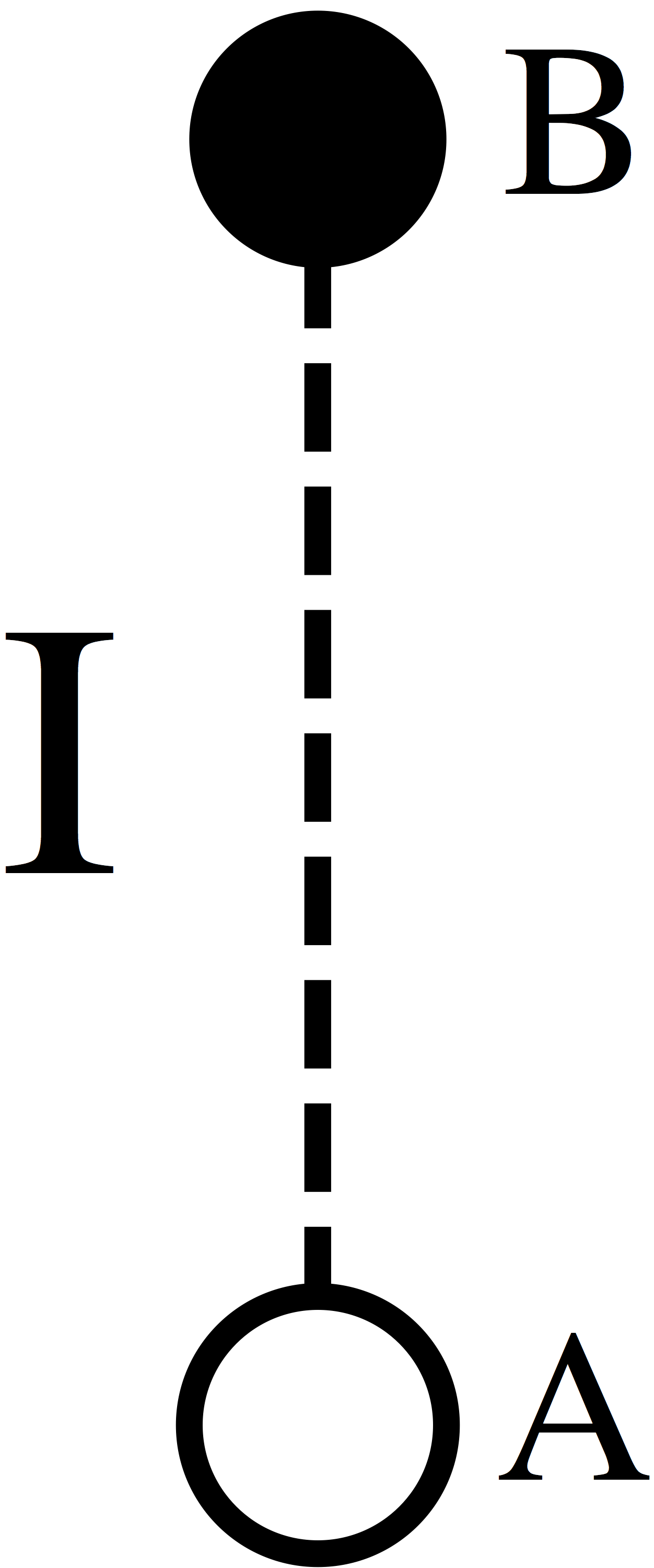}}

\\  
\hline 

$Q_I \begin{bmatrix} \phi_A\\ \psi_B\end{bmatrix} = 
\begin{bmatrix} i \dot{\psi}_B\\ \phi_A\end{bmatrix}
$ 
& \raisebox{-0.5\totalheight}{\includegraphics[width=0.05\textwidth]{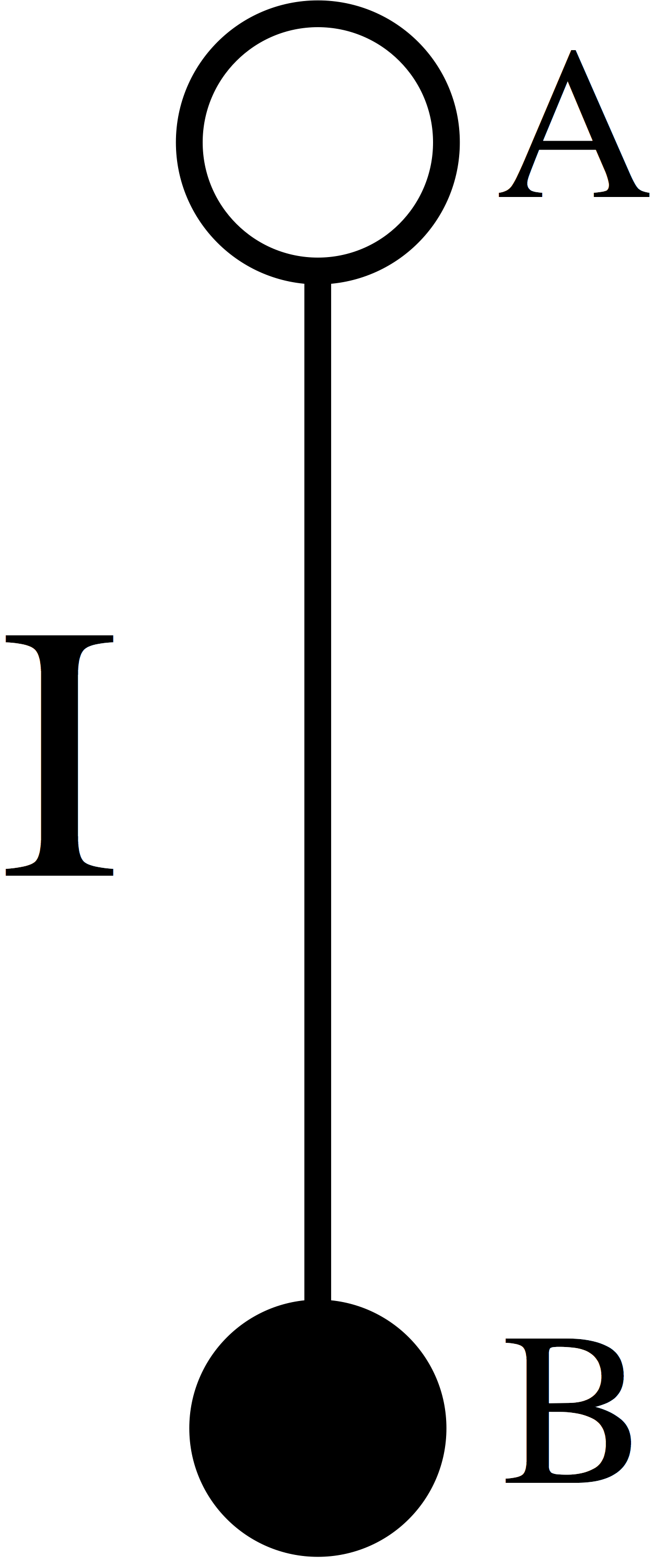}}

& $Q_I \begin{bmatrix} \phi_A\\ \psi_B\end{bmatrix} = 
\begin{bmatrix} -i \dot{\psi}_B\\ -\phi_A\end{bmatrix}
$

& \raisebox{-0.5\totalheight}{\includegraphics[width=0.05\textwidth]{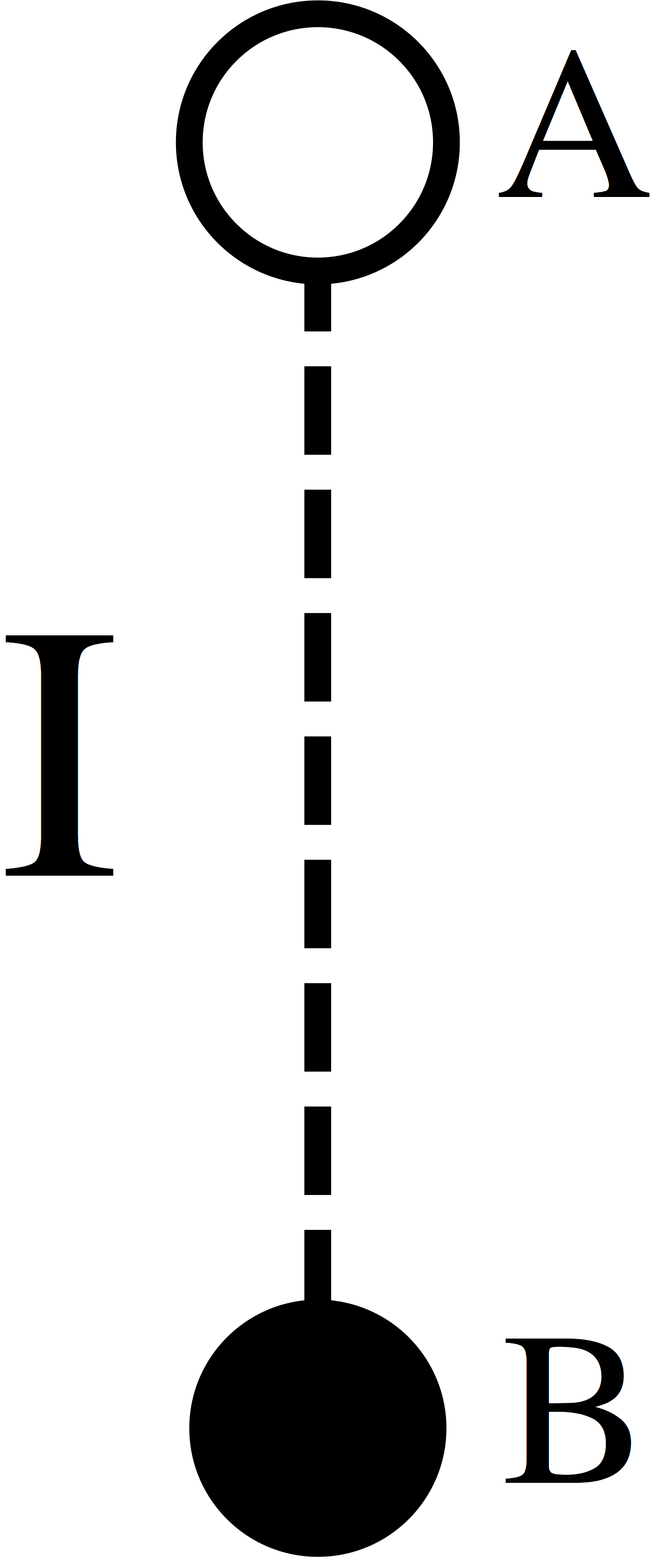}}
\\
 
\end{tabular}
\end{center}
\smallskip

We refer the reader to
\cite{FaGa}, and \cite{Doran1}, \cite{Doran2}, \cite{DFGHIL}, \cite{DILM} for more details on
the correspondence between Adinkras and one-dimensional superalgebras. 

\smallskip
\subsubsection{Adinkras from Codes} 

\begin{figure}[h]
\centering
\includegraphics[width=.4\linewidth]{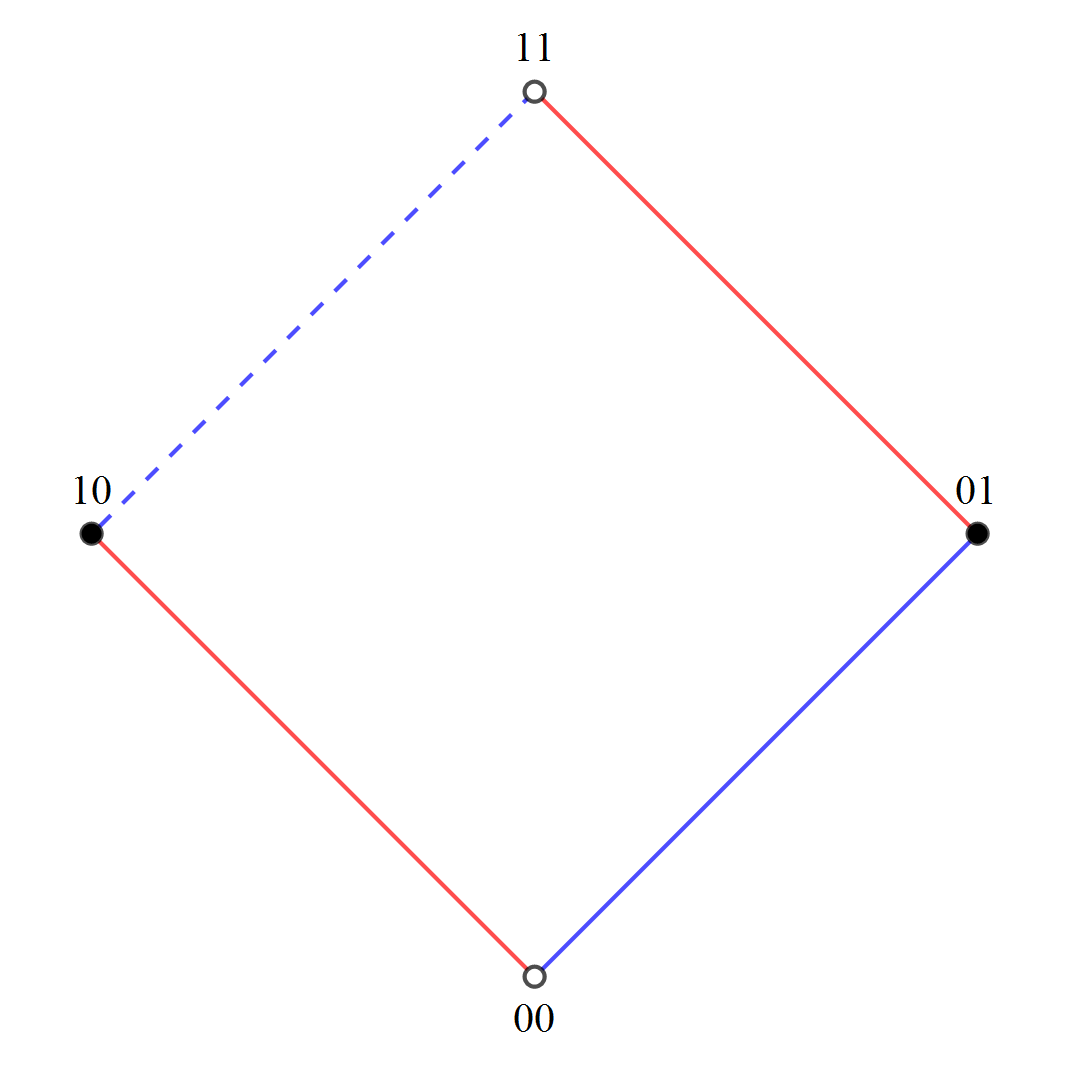}
\caption{\textit{2-cube Adinkra.}\label{fig:2-cube-dashed}}
\end{figure}

The simplest example of an Adinkra is provided by the $N$-cube $A_N$, whose vertices, identified
with binary words of length $N$, one can think of as the Hamming cube. After labeling the $2^N$
vertices with the corresponding binary words, one connects with an edge a pair of vertices whose
words differ by a single entry (words with Hamming distance $1$). The resulting graph can be
colored by assigning color $i$ to an edge connecting vertices whose words differ at the $i$-th 
binary digit. A ranking is obtained by defining $h: V(A) \rightarrow \mathbb{Z}$ as 
$h(v) =$ \textit{\# of 1's in v}. The bipartion into bosons and fermions is obtained by
separating vertices with even ranking (bosons) and odd ranking (fermions). There
are then $2^{2^N - 1}$ possible choices of dashings. An example of a well-dashed 2-cube
is given in Figure~\ref{fig:2-cube-dashed}.

\smallskip

It was shown in Theorem~4.4 of \cite{Doran1} (see also \cite{Doran2})
that all Adinkraizable chromotopologies can be obtained from the cube 
Adinkras via linear codes. Theorem~4.5 of \cite{Zhang} gives a slightly 
more general version of this result that includes multigraphs. More precisely, a
linear binary code $L$ of dimension $k$ is a $k$-dimensional linear subspace 
of the $\Z/2\Z$-vector space $(\Z/2\Z)^N$. The \textit{weight} $wt(c)$ of
codewords $c\in L$ counts the number of $1$'s in the codeword. A code $L$
is \textit{even} if every $c \in L$ has even weight and \textit{doubly-even} 
if the weight of every codeword is divisible by $4$. One can associate a 
chromotopology $A = A_N / L$ to the quotient space $(\Z/2\Z)^N/L$ by 
taking as set of vertices $V(A)$ the equivalence class of vertices, with
an edge of color $i$ connecting two classes $\left[ v \right]$ and $\left[ w \right] \in V(A)$ 
whenever there is at least one edge of color $i$ between a vertex 
$v' \in \left[ v \right]$ and a vertex $w' \in \left[ w \right]$. The properties of
the code $L$ determine properties of the resulting chromotopology $A = A_N / L$.
In particular, the following properties are significant to the construction of Adinkras:
\begin{itemize}
			\item the graph $A$ has a loop iff L contains a codeword of weight 1.
			\item the graph $A$ has a double edge iff L contains a codeword of weight 2.
			\item the graph $A$ can be ranked iff $A$ is bipartite, which is true iff $L$ is an even code.
			\item the graph $A$ can be well-dashed iff $L$ is a doubly-even code. 
\end{itemize}
The main result of Theorem~4.4 of \cite{Doran1}  then shows that Adinkraizable chromotopologies 
are exactly quotients $A = A_N / L$, with $L$ is a doubly-even linear code.

\smallskip

One often denotes by $A_{N,k} = A_N / L$ an Adinkra obtained from a code with $k = \dim(L)$.
		
\bigskip
\section{Adinkras, Dessins and Origami curves}

In the recent paper \cite{DILM}, the authors uncovered a surprising connection between Adinkras
and Grothendieck's theory of dessins d'enfants by showing that the data of an Adinkra diagram 
determine a Belyi pair. 

\smallskip

According to Belyi's theorem \cite{Belyi}, a smooth projective algebraic curve $X$ is 
defined over a number field
if and only if it admits a Belyi map, that is, a branched cover $f: X \to \widehat{\mathbb{C}}=\P^1(\C)$
that is ramified only at the points $0$, $1$, and $\infty$. Moreover, let $X$ be a smooth algebraic curve
defined over a number field and let $f:X \to \widehat\C$ be a Belyi function. Denote by 
$p,q,r$, positive common multiples of the orders of ramifications of $f$ at $0,1,\infty$, respectively,
so that $p^{-1}+q^{-1}+r^{-1}<1$. Let $\Delta=\Delta_{p,q,r}$ 
be the Fuchsian triangle group of signature $(p,q,r)$.
Then it is known (see \cite{Cohen}) that there exists a uniformization $\Phi: \H \to 
X=H \backslash \H$, where $\H$ is the hyperbolic upper half plane, 
ramified at $f^{-1}\{ 0,1,\infty\}$, and where the uniformizing group
is a finite index subgroup $H \subset \Delta$.  The Belyi function then
gives a branched covering $f: X=H \backslash \H \to 
\widehat\C=\Delta_{p,q,r}\backslash \H$. If $p,q,r$ are the ramification
numbers at every point of $f^{-1}(0)$, $f^{-1}(1)$, $f^{-1}(\infty)$, respectively, then
$\Phi$ is the unramified universal cover of $X$. 

\smallskip
\subsection{Dessins d'enfant}

Let $X$ be a compact Riemann surface and let $\widehat{\mathbb{C}}=\P^1(\C)$ denote 
the Riemann sphere. A Belyi map $f: X \rightarrow \widehat{\mathbb{C}}$ is a meromorphic function,  unramified outside of the set $\{0, 1, \infty\}$. A \textit{dessin d'enfant} (sometimes simply referred to  as
\textit{dessin}) is a bipartite graph $\Gamma$ embedded on the surface $X$, obtained by placing white vertices at the points $f^{-1}(1)$, black vertices at the points $f^{-1}(0)$, and curvilinear edges along the preimage $f^{-1}\left( \cI^o \right)$ of the open interval $\cI^o=(0,1)$.
A Belyi pair $(X, f)$ is a Riemann surface $X$ equipped with a Belyi map $f$. It should be noted that every Belyi pair defines a dessin and every dessin defines a Belyi pair.

\begin{ex}\label{ex:Dessin}{\rm
Consider the function $f(x) =  -\frac{(x - 1)^{3}(x - 9)}{64x}$. Figure~\ref{fig:Dessin} shows the associated
dessin.
\begin{figure}[h]
			\begin{center}
				\includegraphics[scale=0.5]{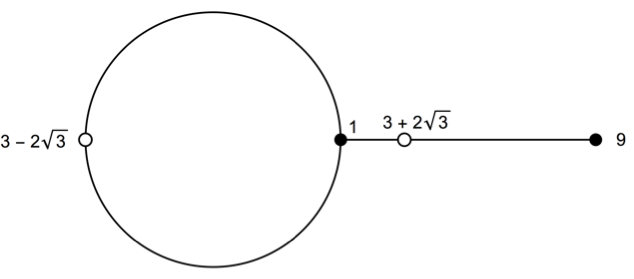}
				\caption{\textit{The dessin corresponding to the equation $f(x) =  -\frac{(x - 1)^{3}(x - 9)}{64x}$ }}
				\label{fig:Dessin}
			\end{center}	
			\end{figure}	}
\end{ex}

It is proved in \cite{DILM} that, given an Adinkra $A_{N,k}$, one can associate to it a dessin.
This is achieved by constructing an embedding of $A_{N,k}$ on a compact Riemann surface 
$X_{N,k}$, of genus 
$g = 1 + 2^{N - k - 3}(N - 4)$ for $N \geq 2$ and $g = 0$ for $N < 2$. The surface $X_{N,k}$
is constructed by
attaching $2$-cells to all consecutively colored $4$-cycles of the Adinkra. Moreover, it
is shown in \cite{DILM} how to obtain the Belyi pair $f: X_{N,k} \to \hat\C$ with the
Adinkra as dessin graph.

It was shown in \cite{ShaVo} that a smooth projective algebraic curve
is defined over a number fields if and only if, as a compact Riemann
surface, it can be triangulated by equilateral triangles. More precisely,
such a triangulation consists of an even number of open cells $T^\pm_j$
homeomorphic to Euclidean triangles. The triangulation of $X$ lifts to a 
triangulation of the universal cover $\H$, such that, for each triangle 
$\tilde T$ in $\H$ there are reflections $\sigma_1,\sigma_2,\sigma_3$ in
the sides of $\tilde T$ that are anti-holomorphic
homeomorphisms of $\H$ preserving the
triangulation, mapping the triangle $\tilde T$ to its neighbor sharing the
side around which the reflection happens, and switching orientation.
This is called a {\em covered symmetric triangulation} (see \cite{Cohen}). 
Given a Belyi map $f: X \to \widehat\C$, a covered symmetric triangulation
is obtained by considering the two hemispheres $\H^\pm$ with
$\H^+\cup \H^-=\P^1(\C) \smallsetminus \P^1(\R)$ and the triangles $T^\pm_j$
in $X$ given, respectively, by the connected components of $f^{-1}(\H^\pm)$.

\smallskip

\begin{lem}\label{ADessinT} For $N>4$, 
an Adinkra $A_{N,k}$ determines a covered symmetric triangulation on
the Riemann surface $X_{N,k}$ with Fuchsian triangle group $\Delta=\Delta_{N,N,2}$.
\end{lem}

\proof 
As shown in \cite{DILM} the Belyi map $f: X_{N,k}\to \widehat\C$ has ramification
of order $N$ at $0$ and $1$ and ramification of order $2$ at $\infty$, hence according
to \cite{Cohen} the Fuchsian triangle group associated to the dessin is $\Delta_{N,N,2}$.
The condition $N>4$ ensures the hyperbolic condition $\frac{2}{N} +\frac{1}{2} <1$ for the
triangle Fuchsian groups. The remaining cases with $N\leq 4$ correspond to genus zero
and genus one surfaces. 
The preimages $f^{-1}(\H^\pm)$ under the Belyi maps constitute the triangles $T^\pm_j$
of the covered symmetric triangulation. In $X_{N,k}$ the pairs of triangles $T^+_j\cup T^-_j$ 
constitute the $2$-cells attached to $A_{N,k}$ along consecutively colored $4$-cycles
in the Adinkra: the diagonal along which the two triangles are joined lies in the preimage
under the Belyi map $f$ of the two arcs $(1,\infty)$ and $(\infty,0)$ in 
$\P^1(\R)=\P^1(\C)\smallsetminus (\H^+\cup \H^-)$, with the preimage of the point $\infty$
lying in the middle of the $2$-cell. 
\endproof

\smallskip
\subsection{Origami curves}
In addition to Grothendieck's theory of dessins d'enfant, another construction
of algebraic curves based on combinatorial data has received considerable
attention: the theory of origami curves. These consist of branched coverings
$p: Y \to E$ ramified only over $\{ \infty \}$.
Origami curves determine Teichm\"uller embeddings of the hyperbolic plane
in the Teichm\"uller space $T_{g,n}$ for given genus $g$ with $n$ marked points. These
are complex geodesics (Teichm\"uller disks). Origami curves have been
largely studied in connection to Teichm\"uller curves and Veech groups, see
\cite{HeSch}, \cite{Moll}.
Moreover, both dessins d'enfant and origami curves have been studied extensively as
possible combinatorial objects carrying an action of the absolute Galois group, with
the goal of gaining a better understanding of the absolute Galois group through
combinatorial models of its action. In particular, various relations between
dessins and origami have been studied, see \cite{HeSch}, \cite{Moll}. For our
purposes, we will especially focus on a construction, given in \cite{Moll} that
associates an origami curve to a dessin d'enfant. 

\smallskip

Let $E$ be an elliptic curve and let $X$ be a compact Riemann surface. 
An \textit{origami} is a finite covering map $p: X \rightarrow E$ ramified 
over a single point $\infty \in E$.

\smallskip

\begin{figure}[h]
	\begin{center}
		\includegraphics[scale=0.4]{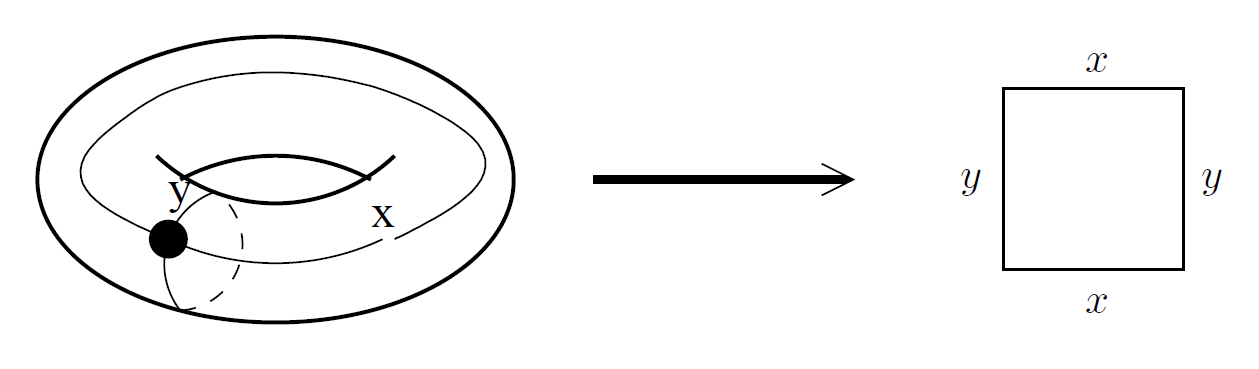}
		\caption{\textit{The torus with generators $x, y$. Reproduced from \cite{HeSch}}}
		\label{fig:torusXY}
	\end{center}	
	\end{figure}

\begin{figure}[h]
		\centering
		\begin{minipage}{.5\textwidth}
			\centering
			\includegraphics[width=.7\linewidth]{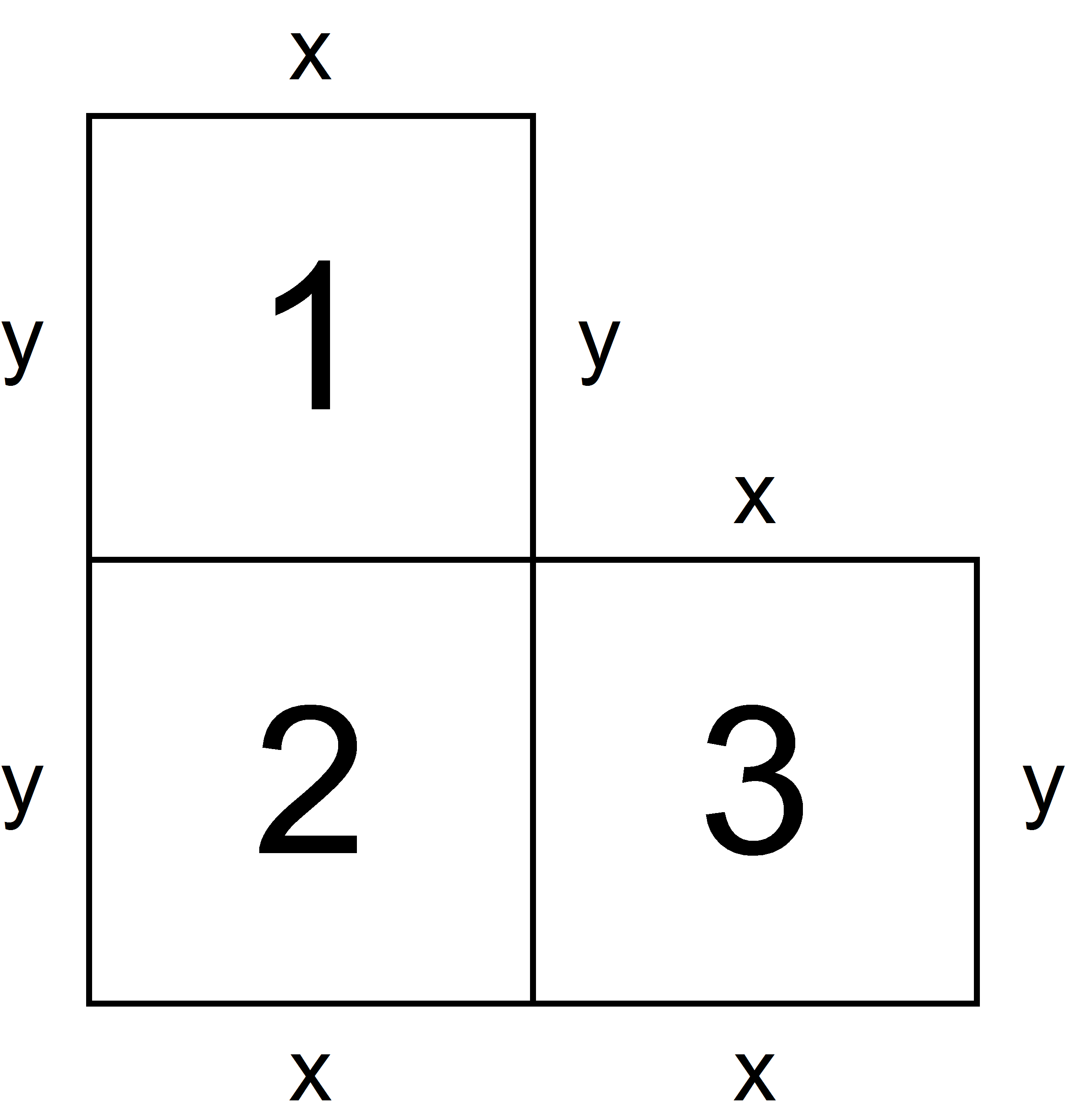}
			\captionsetup{width=0.8\textwidth}
			\captionof{figure}{\textit{A simple example of an origami with opposite edges glued together.}}
			\label{fig:origamiL3Squares}
		\end{minipage}%
		\begin{minipage}{.5\textwidth}
			\centering
			\includegraphics[width=.7\linewidth]{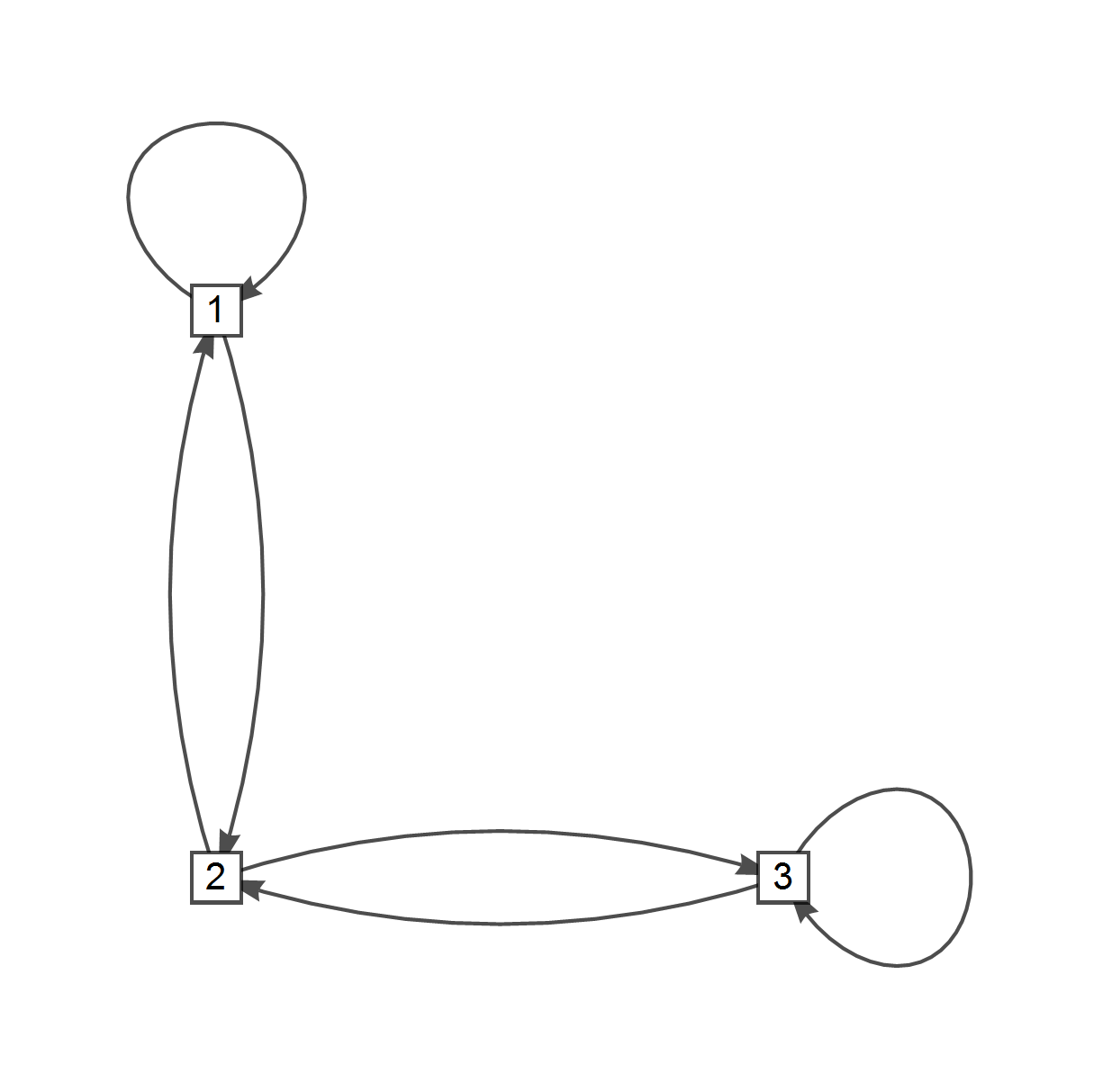}
			\captionsetup{width=0.8\textwidth}
			\captionof{figure}{\textit{The oriented graph descrption of the origami in Figure \ref{fig:origamiL3Squares}}}
			\label{fig:origamiL3}
		\end{minipage}
		\end{figure}

Consider the cell structure of $E$ given by the generators $x, y$ shown in Figure~\ref{fig:torusXY}. Now consider the preimage $p^{-1}(E)$. This creates a finite cell structure on $X$ obtained by gluing cells 
together, as in Figure~\ref{fig:origamiL3Squares}. 

\smallskip

Proposition~5.1 of \cite{HeSch} gives the following equivalent description of origami curves.
An origami $p: X \rightarrow E$ is up to equivalence uniquely determined by 
		\begin{enumerate}
			\item A surface obtained from gluing Euclidean unit squares such that: 
				\begin{itemize}
					\item each left edge is uniquely glued to a right edge and vice versa.
					\item each upper edge is glued to a unique lower one and vice versa.
					\item the result is a connected surface $X$.
				\end{itemize}
				
			\item A finite oriented graph with edges labelled by $x$ and $y$, with the property that each vertex 
			has exactly two incoming edges and two outgoing edges, with one of which type labelled, respectively, 
			by $x$ and $y$. 
				
			\item A monodromy map $\alpha : F_2 \rightarrow S_d$, up to conjugation in $S_d$, where $F_2$ is the free group with 2 generators and $S_d$ is the permutation group of $d$ elements.
			
			\item A finite subgroup $U$ of $F_2$ up to conjugation in $F_2$.
			
		\end{enumerate}

\smallskip
\subsection{Dessins and Origami curves}

We recall here a way of passing from a dessin (a Belyi map $f: X \to \widehat\C$)
to an origami curve, as introduced by 
\cite{Moll} (see also \S 4 of \cite{Nisba}). This construction associates to a Belyi map $f: X \to \widehat\C$ 
an origami $p: Y \to E$, where the curve $Y$ is obtained by first taking the fibered product
$\tilde Y =E \times_{\widehat\C} X$, fibered with respect to the map $f:X \to \widehat\C$ and the quotient
$h: E \to \widehat\C$ by the elliptic involution, 
\begin{equation}\label{tildeXfib}
\tilde Y =E \times_{\widehat\C} X= \{ (z,w) \in E\times X\,:\, h(z)=f(w) \}.
\end{equation}
The map $h: E \to \widehat\C$ is a double cover, ramified at four points $\{ 0,1,\infty,\lambda \}$.
The curve $\tilde Y$ has singularities at those points $(z,w)$ such that $h$ is ramified at $z$
and $f$ is ramified at $w$. The desingularization $Y \to \tilde Y$ is connected and endowed with
a map, which we still denote by $h$, to the elliptic curve, $h: Y \to E$, which is a branched
covering, branched at $\{ 0,1,\infty,\lambda \}$.
Let $m_2: E\to E$ be the multiplication by $2$, the unramified covering that 
maps four Weierstrass points (which we identify with $\{ 0,1,\infty,\lambda \}$) to $\infty$.
Then the composition $p=m_2\circ h: Y \to E$ is ramified at only the point $\infty$ and is
therefore an origami curve. One refers to the origami curve obtained in this way as
the M-origami $O(f)=(p: Y\to E)$ associated to the dessin $f:X \to \widehat\C$.

\smallskip
\subsection{$4$-colored Adinkras and elliptic curves}\label{4colSec}

By the general result of \cite{DILM}, all $A_{N,k}$ Adinkras give rise to dessins, which
determine a Riemann surface $X_{N,k}$ and a Belyi map $f: X_{N,k}\to \widehat\C$,
so that $A_{N,k}$ embeds in $X_{N,k}$ as the dessin obtained as the preimage 
$f^{-1}(\cI)$ with $\cI=[0,1]\subset \widehat\C$. 

\smallskip

\begin{lem}\label{4color}
In the case of Adinkras $A_{N,k}$ with $N=4$ the Riemann surface $X_{4,k}$ is an elliptic curve.
\end{lem}

\proof For $N=4$, the genus formula $g=1+ 2^{N-k-3}(N-4)$ for $X_{N,k}$ immediately implies that
$X_{4,k}$ is an elliptic curve. 
\endproof

\begin{figure}[h]
			\centering
			\begin{minipage}{.5\textwidth}
				\centering
				\includegraphics[width=.7\linewidth]{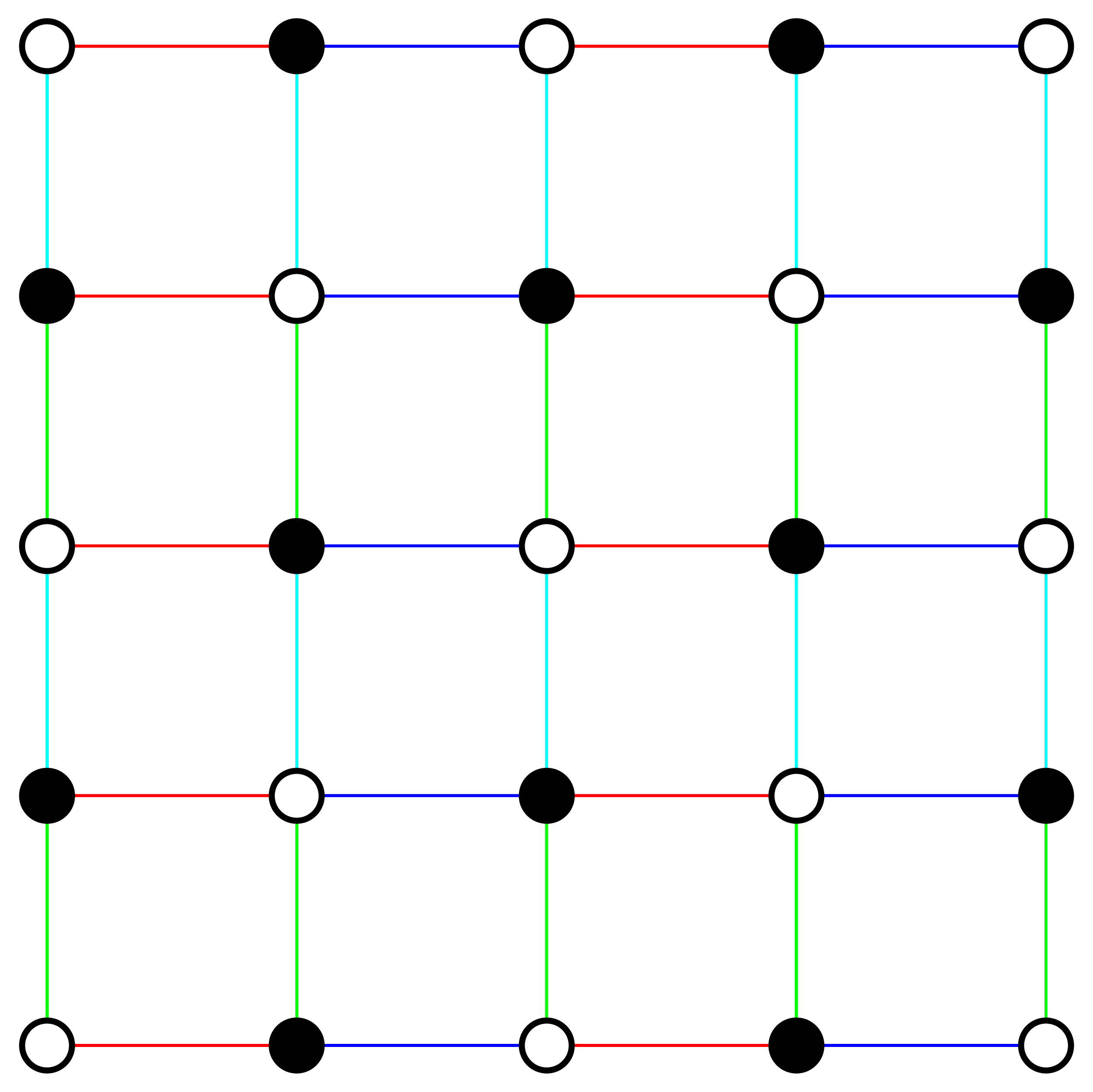}
				\captionsetup{width=0.8\textwidth}
				\captionof{figure}{\textit{The Adinkra $A_{4,0}$ embedded in an elliptic curve, 
				obtained by identifying the top row of edges with the bottom row, and the left 
				column of edges with the right column.}}
				\label{fig:A40origami}
			\end{minipage}%
			\begin{minipage}{.5\textwidth}
				\centering
				\includegraphics[width=.7\linewidth]{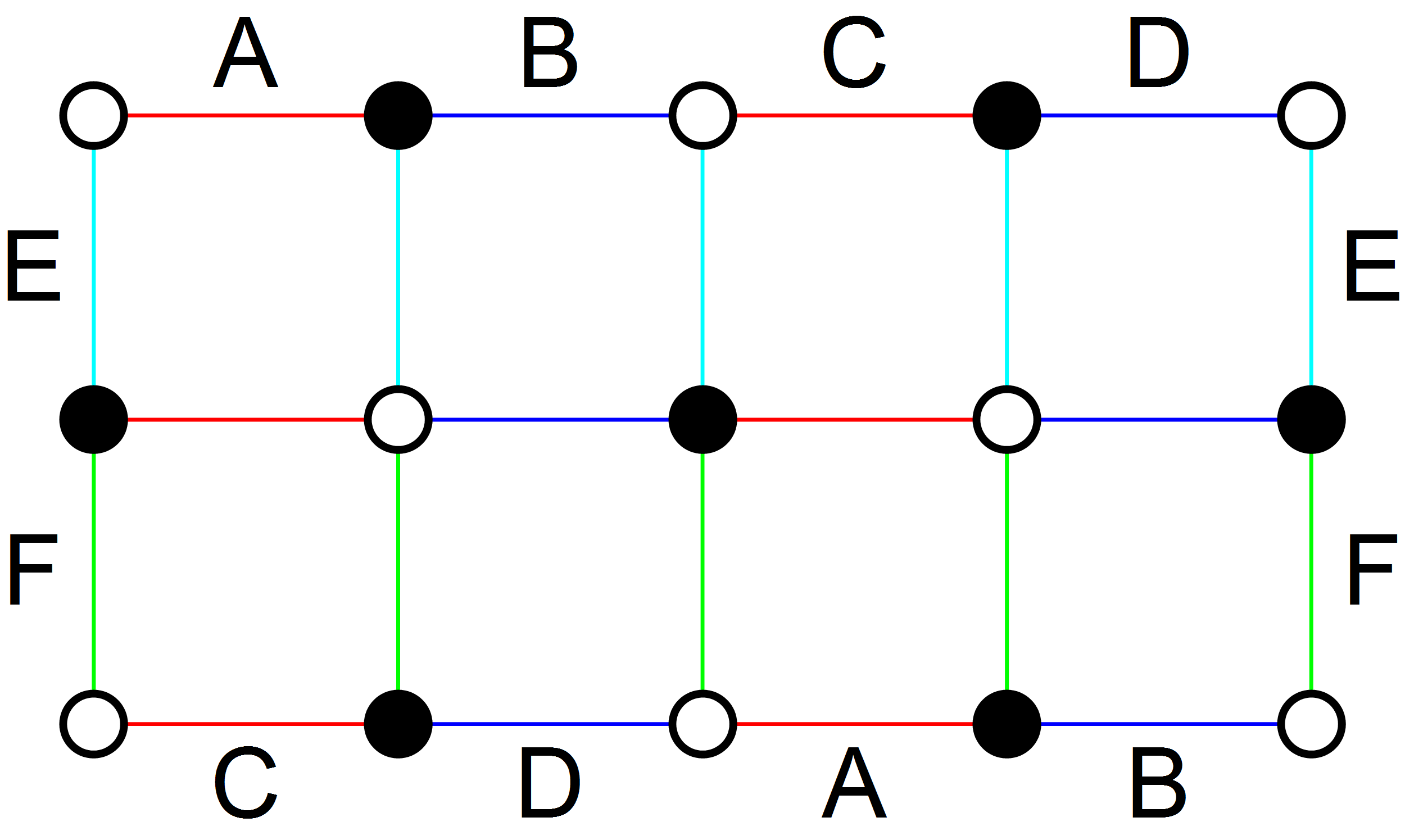}
				\captionsetup{width=0.8\textwidth}
				\captionof{figure}{\textit{The Adinkra $A_{4,1}$ embedded in an elliptic curve, 
				obtained by identifying the edges as labeled.}}
				\label{fig:A41origami}
			\end{minipage}
		\end{figure}

We assign the colors bijectively to the directed edges. Because each vertex has four edges going to unique vertices, this translates to each face being adjacent to four unique faces in the origami picture, resulting in a grid cell structure. This corresponds exactly to the cell structure we would get if we glued $2$-cells to consecutively colored $2$-colored $4$-cycles, as is done in \cite{DILM} to determine the dessin. The choice of consecutively colored $2$-cycles gives us an orientation, just as the directed edges do. 
By identifying edges according to labels we obtain a quotient $E$ that is still topologically a $2$-torus, see
Figures~\ref{fig:A40origami} and \ref{fig:A41origami}, depicting the cases of the Adinkras
$A_{4,0}$ and $A_{4,1}$, respectively.

\medskip
\subsection{Dual graphs of Adinkras and origami}

We present here a relation between Adinkras and
origami curves that uses the characterization recalled above (Proposition 5.1 of 
\cite{HeSch}) of origami in terms of oriented graphs with edges labeled $x$
or $y$, two incoming and two outgoing edges at each vertex, with one of each
type labelled by $x$ and $y$, respectively. We refer to an oriented graph satisfying
these property as an ``origami graph".

\begin{prop} For $N>2$, 
let $A = A_{N,k}$ be an Adinkra, embedded on the Riemann surface $X_{N,k}$,
formed by attaching $2$-cells to all consecutively colored $4$-cycles in $A$. Let
$A^*$ denote the dual graph of $A$, with respect to the given embedding $A\hookrightarrow X_{N,k}$. 
If $N$ is even, then there exists an edge labeling and an orientation such that $A^*$ is an origami graph.
\end{prop}

\begin{figure}[h]
		\centering
	
		\includegraphics[width=\linewidth]{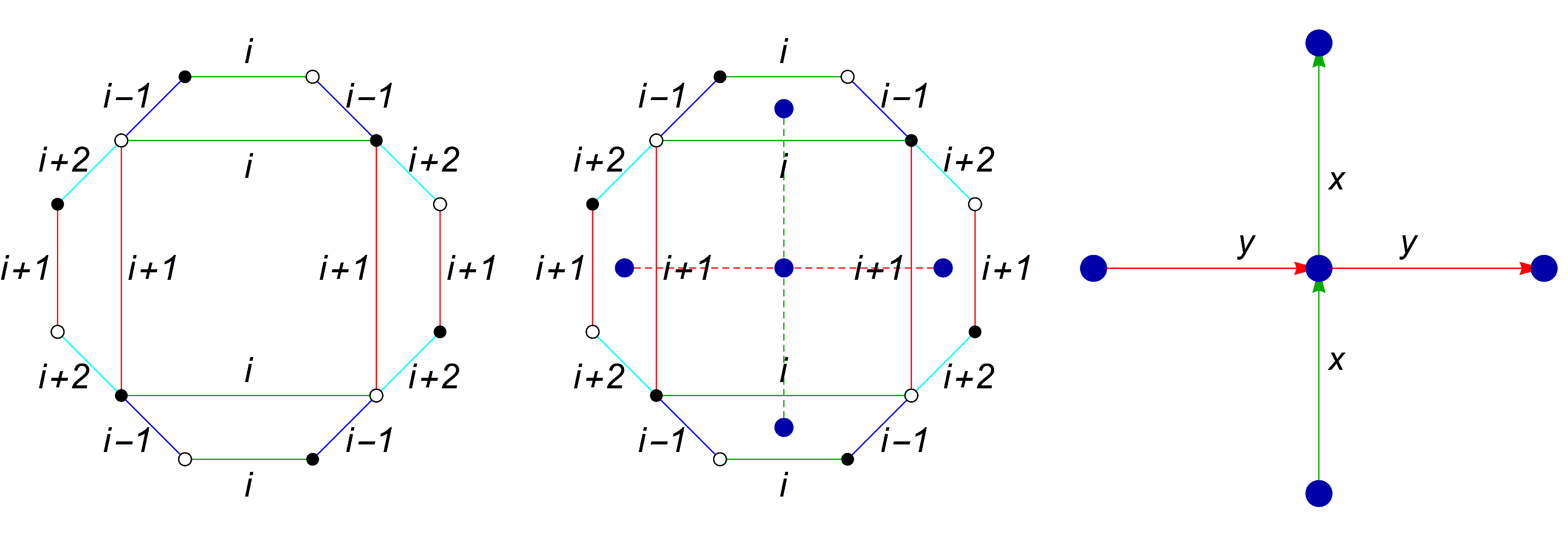}
		\caption{The process of defining the origami. \textit{Left:} the Adinkra $A$ with face $F_{i, i+1}$ adjacent to faces with edges $i-1$ and $i+2$. \textit{Middle:} The Adinkra, $A$, with its dual graph, $A^*$. \textit{Right:} The oriented dual graph, $A^*$, with odd edges labeled $x$ and even edges labeled $y$. The resulting orientation and labeling is consistent with that of an origami. \label{DualFig}}
		
		\end{figure}

\proof Let $E$ and $V$ denote, respectively, the edge and vertex set of $A$.
Similarly, let $E^*$ and $V^*$ denote the edge and vertex set of the dual graph $A^*$.
First observe that, because every face of the embedding $A \hookrightarrow X_{N,k}$ is comprised 
of a consecutively colored $4$-cycle, every vertex in $V^*$ has degree four. 
Let $F_{i, i+1}$ be a face in the embedding $A \hookrightarrow X_{N,k}$ with boundary a consecutively colored
$4$-cycle with colors $\{ i, i+1 \}$. Then each edge with color $i$ in $\partial F_{i,i+1}$ is also incident to a 
face $F_{i-1,i}$ with colors $\{ i-1, i \}$. Clearly these faces exist, as each vertex in $V$ is incident to exactly 
one edge of each color, and these faces are distinct, because if they were the same face, then there would 
exist a $2$-cycle between the colors $\{ i-1, i\}$ which is disallowed by the chromotopology rules. 
By the same argument, each edge of color $i+1$ in $\partial F_{i, i+1}$ is also incident to a face $F_{i+1,i+2}$
with boundary a $4$-cycles with colors $\{ i+1, i+2 \}$. 
Notice also that no other faces can be adjacent to $F_{i,i+1}$ because if there were other such faces $F_{i,j}$, 
then either there would be two edges of the same color incident to a vertex, or there would be a face formed 
by two non-consecutive colors. Neither of these possibilities is allowed by hypothesis. Thus, each face $F_{i,i+1}$ 
is incident to exactly four other faces. Thus, the dual graph $A^*$ is $4$-regular. 
Now let us label the dual graph edges as follows. An edge $e^*\in E^*$ adjacent to a given vertex $v^* \in V^*$
crosses exactly one edge $e$ of the face $F_{i,i+1}$ of the embedding $A \hookrightarrow X_{N,k}$ 
that determines the vertex $v^*=v^*(F_{i,i+1})$ of the dual graph. We label the edge $e^*$ with $x$ or $y$,
according to whether the edge $e$ it crosses is colored by
an odd color (i.e.~$i = 1, 3, 5 ...$), or an even color, respectively.
Since the edges of the face of $A$ are consecutively colored, there are exactly two $x$-labelled edges 
and two $y$-labelled edges adjacent to each vertex $v^*\in V^*$. 
We need the assumption that $N$ is even: if $N$ were odd, then the edges adjacent to 
vertices in $V^*$ corresponding to faces $F_{1,N}$ with colors $\{ 1, N \}$ would all be labeled by $x$.
To show that $A^*$ is an origami graph, we still need to assign an orientation to the graph, in such a way that 
at each vertex one incoming and one outgoing edge are labeled, respectively, by $x$ and $y$. 
Pick an arbitrary vertex and assign a valid orientation to each edge  (with one incoming and outgoing 
edge labelled $x$ and $y$, respectively). Proceed inductively by following the outgoing edge of a given 
label: at the next vertex, by our construction, we know there is another edge labelled $y$, which we
can orient outwards. The case of the $x$-labelled
edges is analogous.  This process terminates once a cycle has been created. Then proceed to the next 
undirected vertex. Note that there cannot exist two conflicting cycles, as this would imply that a vertex 
in $V^*$ has more than two edges of a given label, resulting in a contradiction. The construction
of the origami associated to the dual graph is illustrated in Figure~\ref{DualFig}.
\endproof
		
This construction generalizes the case of the $N=4$ Adinkras described in the previous
Section~\ref{4colSec}, which corresponds to the self-dual graphs. In the rest of this section
we describe a different construction that associates origami curves to Adinkras.

\smallskip
\subsection{Dimers and spin curves}

The starting point for the construction of spectral triples associated to
$1D$-supersymmetry algebras is the work of Cimasoni and Reshetikhin \cite{CiRe}
on dimers on surface graphs and spin structures.

\smallskip

A {\em perfect matching} on a bipartite graph $A$ is a collection of 
edges such that every vertex is incident to exactly one edge of the collection.
Perfect matchings are also referred to as {\em dimer configurations}.
As observed in \cite{DILM}, if $A$ is an Adinkra, then taking the set of
edges of a fixed color in $A$ determines a dimer configuration. 

\smallskip

Let $A$ be a graph embedded on a compact Riemann surface $X$.
A {\em  Kasteleyn orientation} is an orientation of the edges of $A$ with the
property that, when going around the boundary of a face counterclockwise, 
one goes against the orientation of an {\em odd} number of edges. 
As observed in \cite{DILM}, an {\em odd dashing} of an Adinkra $A_{N,k}$
determines a Kasteleyn orientation, when viewing the Adinkras as
embedded in a Riemann surface, $A_{N,k}\hookrightarrow X_{N,k}$,
as recalled above. 

\smallskip

A {\em vertex change} operation on an Adinkra is a transformation that
interchanges the dash/solid coloring of each edge incident to a chosen vertex.
Two choices of dashing on an Adinkra are {\em equivalent} if they can be
obtained from one another by performing a sequence of vertex changes.

\smallskip 

It is shown in \cite{CiRe} (see also \cite{DILM}) that
a {\em dimer configuration} on an embedded graph $A$ on a Riemann surface $X$
determines an isomorphism between {\em Kasteleyn orientations} on $A$ (up to equivalence) 
and {\em spin structures} on $X$.  Equivalent dashings
give equivalent orientations, which correspond to the same spin 
structure on $X$.

\smallskip
\subsection{Products and fibered products}

In addition to the Adinkras considered above, one can include other
combinatorial objects obtained from Adinkras via various types of
operations on graphs. For instance, a Cartesian product (with
additional structure that prescribes coloring, ranking, and dashing) 
was considered in \cite{GaHub}, \cite{Hub} and \cite{IgZha}, 
for $2D$-supersymmetry algebras. These products of Adinkras are
defined in the following way.
Let $A_1 = A_{N_1,k_1}$ and $A_2 = A_{N_2,k_2}$ be two 
Adinkras with colors $1, 2, \dots, N_1$ and $N_1 + 1, N_1 + 2, \dots, N_1 + N_2$, 
respectively. The product $A_1\times A_2$ is defined as follows.
Let $u, v \in V(A_1)$ and $u', v' \in V(A_2)$. We denote the presence of an edge between 
vertices $v_1, v_2$ with the notation $v_1 \sim v_2$. We also 
denote the color of a vertex by $c_{i}(v) \in \{-1, 1\}$, where $1$ is white 
and $-1$ is black. For $(u, u'), (v, v') \in V(A_1 \times A_2)$ we have (see \cite{IgZha}):
		\begin{itemize}
			\item $(u, u') \sim (v, v')$ with color $i$ if and only if: 
				\begin{itemize}
					\item $u \sim v$ with color $i$ {\it and} $u' = v'$
					\item {\it or} $u = v$ {\it and} $u' \sim v'$ with color $i$.
				\end{itemize}
			
			\item  the vertex color of $(u,u')$ is given by $c(u,u') = c_{1}(u)c_{2}(u')$.
			
			\item the ranking of $(u,u')$ is given by $h(u,u') = h_{1}(u) + h_{2}(u')$, where
			$h_i$  are the rankings of each Adinkra.
			
			\item the dashing $d((u, u') \sim (v,v'))$ of edges in the product Adinkra is 
			given by $d_{1}(u \sim u')$, the dashing of $A_1$, or by $d_{2}(v\sim v')+h_1(u)$ mod $2$,
			with $d_2$ the dashing of $A_2$, in the two cases of $(u, u') \sim (v,v')$ with $u'=v'$
			and $u=v$, respectively.
		 \end{itemize}
		 
\smallskip

We consider here a different kind of product of Adinkras, which we view as a fibered product,
because they will correspond to taking fibered products of the corresponding Riemann surfaces
fibered over the Belyi maps.

\begin{defn}\label{fibprodAd}
Let $A_1$ and $A_2$ be Adinkras, with $N_1$ and $N_2$ the respective number of colors. 
Let $f_i: X_i \to \widehat\C$, for $i=1,2$, be the associated Belyi maps. 
The fibered product $A=A_1\times_\cI A_2$, with $\cI=[0,1]$,
is defined as the graph with set of vertices
$V(A)=V_0(A_1)\times V_0(A_2) \cup V_1(A_1)\times V_1(A_2)$, where $V_i(A_j)$
denotes the vertices of $A_j$ that are colored $i$ in the bipartition.  The edges are
given by all pairs of edges $(e_1,e_2)\in E(A_1)\times E(A_2)$ with endpoints in $V(A)$.
These are all edges in $E(A_1)\times E(A_2)$, since the $A_i$ are bipartite. 
\end{defn}

The graph obtained in this way has a natural interpretation in terms of the fibered product
of the assiciated Riemann surfaces, fibered over the Belyi maps.

\begin{lem}\label{Adinkrafib}
Let $A_{N_i,k_i}$, with $i=1,2$ be two Adinkras, and let $f_i: X_{N_i,k_i}\to \widehat\C$ be
the associated Belyi maps. Let $Y$ be the desingularization of the fibered product
$\tilde Y=X_{N_1,k_1}\times_{\widehat\C} X_{N_2,k_2}$ fibered along the Belyi maps,
with $f: Y\to \widehat\C$ the resulting branched cover. 
The graph $A=A_{N_1,k_1}\times_\cI A_{N_2,k_2}$ can be identified with the 
preimage $f^{-1}(\cI)$ in $Y$.
\end{lem}

\proof
The set of vertices $V(A)$ is given by $V(A)=f_1^{-1}(0)\times f_2^{-1}(0)\cup f_1^{-1}(1)\times f_2^{-1}(1)$,
hence it corresponds to the locus $\{ (z,w)\in X_{N_1,k_1}\times X_{N_2,k_2}\,:\, f_1(z)=f_2(w)=0\} \cup
\{  (z,w)\in X_{N_1,k_1}\times X_{N_2,k_2}\,:\, f_1(z)=f_2(w)=1 \}\subset X_{N_1,k_1}\times_{\widehat\C} X_{N_2,k_2}$. The edges $E(A)$ can be identified with the set 
$\{ (z,w)\in X_{N_1,k_1}\times X_{N_2,k_2}\,:\, f_1(z)=f_2(w)\in \cI \}\subset 
X_{N_1,k_1}\times_{\widehat\C} X_{N_2,k_2}$. 
\endproof

\begin{lem}\label{fibprocol}
The fibered product $A=A_1\times_\cI A_2$
is a chromotopology with a
 ranking $h: V(A) \to \Z$, which send
white/black vertices of the bipartition to even/odd numbers.
When the numbers of edge colorings $N_1$ and $N_2$ are coprime,
there is a unique such chromotopology structure, while otherwise
there are several inequivalent structures, corresponding to the
different choices of colorings. One of these corresponds to the
embedding $A=A_{N_1,k_1}\times_\cI A_{N_2,k_2} \simeq f^{-1}(\cI)$ in $Y$,
while the other choices correspond to other surfaces.
\end{lem}

\proof The graph $A$ is bipartite with $V_0(A)=V_0(A_1)\times V_0(A_2)$ and
$V_1(A)=V_1(A_1)\times V_1(A_2)$. The edges can be colored by $N={\rm lcm}\{N_1, N_2\}$ colors, 
with an edge $(e_1,e_2)$ colored by $(c_1(e_1),c_2(e_2)$, with $c_i$ the coloring on $A_i$. 
Note that, when going around a vertex in $X_1$, the colors go cyclically from $1$ to $N_1$, and
similarly on $X_2$, with colors cyclically ordered from $1$ to $N_2$. Thus, on the fibered
product, the elements $c_i$ in the pairs $(c_1,c_2)$ have periodicities $N_i$. This implies that, when 
$N_1$ and $N_2$ are not coprime, the different orbits determine different
choices of rainbows of colors, hence different possible resulting chromotopologies. 
Since the graphs $A_i$ are bipartite, the set of edges in $A$ between two vertices
$v=(v_1,v_2)$ and $w=(w_1,w_2)$ consists of the product $E_{v_1,v_2}(A_1)\times E_{w_1,w_2}(A_2)$
of the sets of edges in $A_i$ connecting the two respective vertices.
Since the $A_i$ have no parallel edges, there is only one edge in each set $E_{v_1,v_2}(A_1)=\{ e_1 \}$
and $E_{w_1,w_2}(A_2)=\{ e_2 \}$, hence there is a unique $e=(e_1,e_2)$ connecting
$v=(v_1,v_2)$ and $w=(w_1,w_2)$ in $A$, with color $c(e)=(c_1(e_1),c_2(e_2))$. Thus, each
vertex in $A$ has valence $N$ and is incident to exactly one edge of each color.
Given two distinct colors $(i,j)\neq (i',j')$, consider the set of edges of $A$ with color either
$c(e)=(i,j)$ or $c(e)=(i',j')$. The condition $(i,j)\neq (i',j')$ occurs when either $i\neq i'$ or
when $i=i'$ and $j\neq j'$. In the first case, the set we are considering consists of
$$ ( c_1^{-1}(i) \cup c_1^{-1}(i') ) \times E(A_2). $$
In this case, the left factor consists of a disjoint union of $4$-cycles, hence the product set
also does, as it is a disjoint union of a copy of such a disjoint union of $4$-cycles, for each
choice of an element of $E(A_2)$. In the second case, the set we are considering consists of
$$ c_1^{-1}(i) \times ( c_2^{-1}(j)\cup c_2^{-1}(j')). $$
Again the second factor is a disjoint union of $4$-cycles, hence the product also is. This shows
that the fibered product $A$ is an $N$-chromotopology. To show that it is an Adinkra, we
need to check that it is well-dashed and it has a ranking $h:V(A)\to \Z$ with $h(V_0(A))\subset 2\Z$
and $h(V_1(A))\subset 2\Z+1$. We obtain a ranking with the desired property by setting
$h(v_1,v_2)=h_1(v_1)\cdot h_2(v_2)$, where $h_i: V(A_i)\to \Z$ are the rankings of the Adinkras $A_i$.
Since $h_1(v_1)$ and $h_2(v_2)$ are either simultaneously even or simultaneously odd 
for $(v_1,v_2)\in V(A)$, their product is, respectively, even on $V_0(A)$ and odd on $V_1(A)$.
\endproof

As an example where the construction above gives rise to different surfaces with different
rainbows of colors, consider the case where both $A_1$ and $A_2$ are isomorphic to the 
same Adinkra with $N_1=N_2=8$, and with $16$ vertices ($8$ bosons and $8$ fermions).
Then the construction of Lemma~\ref{fibprocol} above
determines $8$ different surfaces, each with $8$ colors, corresponding to
the different choices of $c_1-c_2$ mod $8$, rather than a single surface with $64$ different
colors. Indeed, a rainbow of $64$ colors would not be compatible with the Adinkra structure, 
which would require, in that case, at least $2^{32}$ vertices rather than $128$ 
($64$ bosons and $64$ fermions). (This example was suggested to us by Kevin Iga.)

\smallskip

In the rest of this section, when we refer to ``the fibered product chromotopology"
we mean the choice of rainbow of colors in Lemma~\ref{fibprocol}, such that the
resulting chromotopology corresponds to the Riemann surface $Y$ with
Belyi map $f: Y \to \widehat\C$, with the
embedding $A=A_{N_1,k_1}\times_\cI A_{N_2,k_2} \simeq f^{-1}(\cI)$.

\smallskip

The dashings $d_i$ on the Adinkras $A_i$, do not immediately extend to
a dashing on the fibered product Adinkra $A=A_1\times_\cI A_2$ constructed as
above, unlike the case of the Cartesian product structure recalled above, considered 
in \cite{IgZha}. To see where the problem lies, notice that the $2$-colored $4$-cycles
in $A$ correspond to choices of two colors $(i,j)\neq (i',j')$. This corresponds to the
cases $i=i'$ and $j\neq j'$, or $i\neq i'$ and $j\neq j'$, or $i\neq i'$ and $j=j'$. In the
first two cases the dashing $d_1$ would give a dashing on $A$ with the property that
each of these $2$-colored $4$-cycles has an odd number of dashed lines, and
similarly the dashing $d_2$ would work for the second and third case, but each
separately would not cover all cases, as the remaining cases would have an even
number of dashings. 

\smallskip

However, as shown in \cite{DILM}, using the result of \cite{CiRe} on Kasteleyn
orientations and spin curves, the choice of a dashing on an Adinkra $A_{N,k}$ determines
a dimer configuration, hence a spin structure on the curve $X_{N,k}$, in such a way
that equivalent dashings determine the same spin structure. Thus, it is possible to
show that the dashings on the two Adinkras $A_i$ determine a choice of dashing
(up to equilavence) on the fibered product $A=A_1\times_\cI A_2$ by showing that
a choice of spin structures on the respective Riemann surfaces $X_i$ determines
uniquely a choice of a spin structure on the resulting Riemann surface $Y$ obtained
by desingularizing $\tilde Y=X_1\times_{\widehat\C} X_2$. 

\begin{lem}\label{spinfibprod}
There is a one-to-one correspondence between pairs of dashings up to vertex change
equivalence on the Adinkras $A_i$ and dashings on the fibered product
chromotopology $A=A_1\times_{\cI}A_2$ up to vertex change equivalence.
\end{lem}

\proof Because of the one-to-one correspondence between dashings up to
vertex change equivalence on a chromotopology $A$ and spin structures
on the associated Riemann surface $X$ with $A\subset X$, it suffices to
show that there is a one-to-one correspondence between pairs $(\fs_1,\fs_2)$
of spin structures on $X_1$ and $X_2$ and spin structures on the fibered
product $Y$. It is well known that there are exactly $2^{2g}$ different spin
structures on a Riemann surface of genus $g$. Thus, we only need to verify
the relation between the genera $g_1=g(X_1)$ and $g_2=g(X_2)$ and the
genus $g=g(Y)$. Since the base space $\widehat\C$ of the fibered product
is simply connected, the cohomology of the fibered product $\tilde Y=X_1\times_{\widehat\C} X_2$
is computed by the Eilenberg--Moore spectral sequence with $E_2^{p,q}
=Tor^{p,q}_{H^*(\widehat\C)}(H^*(X_1), H^*(X_2))$, converging to 
$H^*(X_1\times_{\widehat\C} X_2)$. This reduces to computing 
$H^*(X_1)\otimes_{H^*(\widehat(\C)} H^*(X_2)$ as a tensor product of graded
modules over a graded ring. This gives $H_1(\tilde Y)\simeq \Z^{2(g_1+g_2)}$,
hence the genera add, $g=g_1+g_2$.
\endproof

\medskip
\subsection{M-origami of Adinkras}

In the rest of this section we show that Adinkra graphs admit embeddings in origami curves.
We use the construction of the Riemann surface $X_{N,k}$ and the Belyi map $f: X_{N,k}\to
\widehat\C$ associated to an Adinkra $A_{N,k}$, as in \cite{DILM}, together with the construction
of M-origami of \cite{Moll}, which obtains an origami curve $O(f)=(p: Y\to E)$ from a Belyi
map $f:X \to \widehat\C$, to show that the Adinkra graph can be embedded (with a choice
of $2^{\# E(A_{N,k}}$ embeddings) in the curve $Y=Y_{N,k}$.

\smallskip

\begin{lem}\label{Aorigami}
Let $A=A_{N,k}$ be an Adinkra. There exists an origami curve $Y_{N,k}$, with
$p: Y_{N,k}\to E$ ramified only at $\{ \infty \}$, determined by the Adinkra, and
a collection of $2^{\# E(A)}$ embeddings of the Adinkra $A_{N,k}$ in $Y_{N,k}$.
\end{lem}

\proof
Let $X_{N,k}$ be the Riemann surface associated to the Adinkra $A=A_{N,k}$, with
Belyi map $f: X_{N,k}\to \widehat\C$. 
Following the construction of the origami curve $O(f)=(p: Y_{N,k} \to E)$ associated to
this Belyi map. The origami curve is determined by the Adinkra $A_{N,k}$ because
the Riemann surface $X_{N,k}$ and the Belyi map $f: X_{N,k}\to \widehat\C$ are. 
Let $\Gamma$ denote the graph in $E$ obtained as $h^{-1}(f(A_{N,k})$,
where $h: E \to \widehat\C$ is the double cover ramified at $\{ 0,1,\infty,\lambda \}$. 
Since the image $f(A_{N,k})$ in $\widehat\C$ consists of the interval $\cI=[0,1]$,
the set of vertices $V(\Gamma)=h^{-1}(0)\cup h^{-1}(1)$ consists of two points,
$V(\Gamma)=\{0,1\}$ and the set of edges consists of two parallel edges 
connecting the two vertices, $E(\Gamma)=\{ e_1,e_2\}=h^{-1}(\cI^o)$, with 
$\cI^o=(0,1)$. Consider then the preimage in $\tilde Y_{N,k}=E\times_{\widehat\C} 
X_{N,k}$ of the interval $\cI$ in $\widehat\C$. This consists of a graph $\tilde A$
with set of vertices $V(\tilde A)=V(A)$ and set of edges $E(\tilde A)=E(A)\cup E(A)$
consisting of two copies of the set of edges of $A$, with each edge of $A$ replaced
by a pair of parallel edges. Thus, we obtain in this way $2^{\# E(A)}$ ways of
embedding the graph $A=A_{N,k}$ in $\tilde Y_{N,k}$ (hence in the desingularization
$Y_{N,k}$) given by choosing, in all possible ways, one of the two parallel edges.
\endproof

\bigskip
\section{Spectral triples from Supersymmetric Algebras}

As shown in \cite{CoS3}, 
in the setting of Noncommutative Geometry, it is possible to
encode and generalize the data of compact smooth Riemannian
spin manifolds in the form of a triple $(\cA,\cH,D)$ of an
involutive algebra, a Hilbert space on which it is represented
by bounded operators, and a self-adjoint Dirac operator on the Hilbert
space, with compact resolvent and with bounded commutators
with elements of the algebra. In the case of an ordinary compact
smooth spin manifold $X$, these data are $(\cC^\infty(X), L^2(X,\bS), \Dirac)$
with $\bS$ the spinor bundle and $\Dirac$ the Dirac operator on $X$,
for a given choice of spin structure. This axiomatization makes it possible
to extend methods of Riemannian geometry to spaces that are not
ordinary manifolds, including quantum groups, fractals, noncommutative
spaces (like noncommutative tori), and the almost-commutative
geometries used for the construction of particle physics models.
The spectral action functional of a spectral triple $(\cA,\cH,D)$ 
is a natural action functional, which corresponds to a theory of
gravity on this geometry. In the case of almost-commutative
geometries ts large energy asymptotic expansion
recovers classical physical action functionals for gravity
coupled to matter. 

\bigskip
\section{Spectral Action}

The spectral action is a natural construction of an action functional for
a spectral triple, defined in terms of the spectrum of the Dirac operator,
in the form of a regularized trace $\Tr(f(D/\Lambda))$, where $f$ is a
suitable test function (a smooth approximation to a cutoff function) and
$\Lambda$ is an energy scale. It was originally introduced in \cite{CC},
and it became an extremely useful tool for particle physics and gravity
models based on noncommutative geometry, see for instance \cite{WvS}
for a general survey. 

\smallskip

Here we associate to a 1D supersymmetry algebra a spectral triple and
a spectral action functional, through the construction of \cite{DILM}
of the Riemann surface $X_{N,k}$ associated to an Adinkra graph $A_{N,k}$.
We show that the spectral action can be computed using the Selberg
trace formula. We then refine the construction to include the Super
Riemann Surface structure on $X_{N,k}$ determined by the Adinkra,
and we relate the resulting super spectral action to the Selberg supertrace
formula of \cite{BarMan}. 

\smallskip

We then also discuss the spectral geometry and spectral action
associated to the origami curve $O(f)=(p:Y_{N,k}\to E)$ associated
to the Belyi map $f: X_{N,k} \to \widehat\C$, as another possible
construction of a spectral geometry associated to a 1D supersymmetry algebra.

\smallskip
\subsection{Selberg trace formula and the Laplacian spectral action}

Let $X =\Gamma\backslash \H$ be a compact hyperbolic Riemann surfaces
of genus $g=g(X)\geq 2$, uniformized by a Fuchsian group $\Gamma \subset \SL_2(\R)$,
endowed with the hyperbolic metric of constant curvature $-1$ induced from the
hyperbolic upper half place $\H$, with $\Gamma$ acting by isometries. 

\begin{defn}\label{DeltaSA}
Let $\{ \lambda_j = \frac{1}{4} +r_j^2 \}_{j\in \Z_+}$ be the spectrum of the Laplacian $\Delta$
in the hyperbolic metric of constant curvature $-1$ on $X$. 
For $f\in \cS(\R)$ a rapidly decaying even test function, the Laplacian spectral action on $X$
is given by 
\begin{equation}\label{LaplaceSA}
\cS_{\Delta,f}(\Lambda):= \sum_{j=0}^\infty f(r_j/\Lambda) .
\end{equation}
\end{defn}

Let $\cG_X$ denote the set of oriented closed geodesics on $X$. For an oriented
closed geodesic $\gamma \in \cG_X$ let $\ell(\gamma)$ denote the length and
$N_\gamma =\exp(\ell(\gamma))$ the norm. Moreover, let $\lambda(\gamma):=\ell(\gamma_0)$ 
where $\gamma_0$ is the unique oriented primitive closed geodesic, such that 
$\gamma =\gamma_0^m$ for some $m\in \N$.
The Laplacian spectral action can be computed via the Selberg trace formula.

\begin{lem}\label{SpActDelta}
Let $h\in\cC^\infty(\R)$ be an even compactly supported test function. We assume
that ${\rm supp}(h)=[-1,1]$. Let $f\in \cS(\R)$ be the rapidly decaying even test function
obtained as Fourier transform $f=\hat h$. The Laplacian spectral action on $X$ satisfies
\begin{equation}\label{SelbLaplaceSA}
\cS_{\Delta,f}(\Lambda) = \Lambda^2 (g(X)-1) \int_0^\infty r f(r) \tanh(\Lambda \pi r)
\, dr + \Lambda \sum_{\gamma \in \cG_{X,\Lambda}} \frac{\lambda(\gamma)}{N_\gamma^{1/2} - N_\gamma^{-1/2}}
h (\Lambda \log N_\gamma),
\end{equation}
where the sum on the right hand side is over the set of oriented closed geodesics 
\begin{equation}\label{GXL}
\cG_{X,\Lambda}=\{ \gamma \in \cG_X\,:\, \ell(\gamma) \leq \frac{1}{\Lambda} \}.
\end{equation}
\end{lem}

\proof The Selberg trace formula gives
\begin{equation}\label{Selb}
\sum_{j=0}^\infty f(r_j) = \frac{A(X)}{4\pi} \int_0^\infty r f(r) \tanh(\pi r)
\, dr + \sum_{\gamma \in \cG_X} \frac{\lambda(\gamma)}{N_\gamma^{1/2} - N_\gamma^{-1/2}}
h (\log N_\gamma), 
\end{equation}
where $A(X)$ is the area of the surface $X$, which by Gauss--Bonnet satisfies 
$A(X) =4\pi (g-1)$, with $g=g(X)\geq 2$ the genus of $X$. The expression
\eqref{SelbLaplaceSA} is then an immediate consequence of \eqref{Selb},
after replacing $f(r)$ with $f_\Lambda(r)=f(r/\Lambda)$ and $h(s)$ with
$h_\Lambda(s)=\Lambda\, h(\Lambda s)$, so that $f_\Lambda =\hat h_\Lambda$.
Since the function $h$ has support the interval $[-1,1]$, the function $h_\Lambda$
has support $[-1/\Lambda,1/\Lambda]$, hence only the geodesics $\gamma$ in
the set $\cG_{X,\Lambda}$ contribute to the sum.
\endproof

\begin{cor}\label{hypSADelta}
Let $\cC_\Gamma$ denote the set of conjugacy classes of primitive simple hyperbolic elements in
the Fuchsian group $\Gamma$. For $P\in \cC_\Gamma$, let $t_P$ denote the translation 
length of an element of $P$.
The Laplacian spectral action on the hyperbolic surface $X=\Gamma\backslash \H$ satisfies
\begin{equation}\label{hypSelbLaplaceSA}
\begin{array}{rl}
\cS_{\Delta,f}(\Lambda) = &
\displaystyle{\Lambda^2 (g(X)-1) \int_0^\infty r f(r) \tanh(\Lambda \pi r)
\, dr} \\[4mm] + & \displaystyle{\Lambda \sum_{P\in \cC_\Gamma} 
\arccosh(\frac{t_P}{2}) \sum_{\ell \in S_\Lambda(P)} 
\frac{h(\Lambda 2\ell \arccosh(\frac{t_P}{2}))}{\sinh(\ell \arccosh(\frac{t_P}{2}))}} ,\end{array}
\end{equation}
where, for a given $P\in \cC_\Gamma$, the sum is over the finite set
\begin{equation}\label{CGL}
S_\Lambda(P)=\{ \ell \in \N \,:\, 2\ell \arccosh(t_P/2) \leq 1/\Lambda \}.
\end{equation}
\end{cor}

\proof
 Since all elements in $\Gamma \smallsetminus \{ 1 \}$ are hyperbolic, it is well known that the
 Selberg trace formula \eqref{Selb} can be rewritten equivalently in terms of conjugacy classes $\cC_\Gamma$  as
\begin{equation}\label{Selb2}
\begin{array}{rl}
\displaystyle{\sum_{j=0}^\infty f(r_j)} = & 
\displaystyle{(g(X)-1) \int_0^\infty r f(r) \tanh(\pi r)
\, dr} \\[4mm] + & \displaystyle{\sum_{P\in \cC_\Gamma} \arccosh(\frac{t_P}{2}) \sum_{\ell=1}^\infty
\frac{h(2\ell \arccosh(\frac{t_P}{2}))}{\sinh(\ell \arccosh(\frac{t_P}{2}))}} . \end{array}
\end{equation}
Replacing $f(r)$ with $f_\Lambda(r)=f(r/\Lambda)$ and $h(s)$ with
$h_\Lambda(s)=\Lambda\, h(\Lambda s)$, we obtain \eqref{hypSelbLaplaceSA}.
Since the function $h$ has support $[-1,1]$, for a given $P\in \cC_\Gamma$, the only
integers $\ell$ contributing to the sum are those in the set $S_\Lambda(P)$. 
\endproof

\smallskip
\subsection{The Dirac spectral action on a compact Riemann surface}

In noncommutative geometry, instead of working with the Laplacian $\Delta$ and the
Laplace spectral action discussed above, one considers the Dirac operator $D$ and the
Dirac Laplacian $D^2$. Indeed, Dirac operators are abstracted to the more general
setting of spectral triples $(\cA,\cH,D)$, which generalize to possibly noncommutative
settings the spin geometry $(\cC^\infty(X), L^2(X,\bS), D)$, with $\bS$ the spinor
bundle of a compact Riemannian spin manifold $X$, see \cite{CoS3} for more details.
The spectral action functional, for a Dirac operator $D$ of a spectral triple, is defined 
(see \cite{CC}), for $f\in \cS(\R)$ an even rapidly decaying test function, as
\begin{equation}\label{SpactDirac}
\cS_{D,f}(\Lambda)  :=  \Tr(f(D/\Lambda)).
\end{equation}

In the case of a compact Riemann surface $X$ of genus $g\geq 2$, with the
constant negative curvature hyperbolic metric, the behavior of the spectral
action functional \eqref{SpactDirac} is similar to the behavior of the
Laplacian spectral action discussed in the previous subsection. The effect on
the Selberg trace formula of replacing the Laplacian $\Delta$ by the Dirac Laplacian 
$D^2$ is discussed in \cite{Bolte}. When we adapt these results to the argument
given in the previous subsection, we obtain the following result for the
spectral action of a hyperbolic surface. 

In the case of the Dirac operator, the choice of the spin structure on the Riemann
surface $X$ is encoded in the choice of the spinor bundle $\bS$ over $X$. A vector
bundle on $X$ corresponds to a trivial pullback bundle on the universal cover,
$\pi^*(\bS)=\H \times S$, with $S$ the fiber of $\bS$,  
together with the datum of a transition function $\sigma: \H \times \Gamma \to \GL(S)$,
where $\Gamma \subset \PSL_2(\R)$ is the uniformizing Fuchsian group, 
so that spinor sections of $\bS$ are identified with functions $\psi: \H \to S$ with
$\psi(\gamma z) =\sigma(z,\gamma)\psi(z)$. The transition function $\sigma$ can
be encoded by the pair of an automorphy factor $j: \H \times \tilde\Gamma \to \GL(S)$
and a character $\chi: \tilde\Gamma \to U(1)$, where $\tilde\Gamma \subset \SL_2(\R)$
is such that $\Gamma =\tilde\Gamma/\{ \pm 1 \}$, with $\chi(-1)=-1$, see \S III of \cite{Bolte}
and \cite{Roelke}, so that the spinor sections satisfy the automorphic condition
$\Psi(\gamma z)=\chi(\gamma) j(z,\gamma) \Psi(z)$, where $\Psi(z)=(\psi_1(z),\psi_2(z))$
and $j(z,\gamma)$ diagonal with entries $j(z,\gamma,\pm 1)$. (We consider here only
the case of the Dirac operator $D=D_1$ of \cite{Bolte} with weight $k=1$, that is, the standard
Dirac operator.) The matching between the eigenfunctions $\Psi=(\psi_1,\psi_2)$ of $D$ with 
eigenvalue $\lambda$ and the eigenfunctions $\psi=\psi_1$ of the Laplacian with
eigenvalue $\lambda^2 +1/4$ is proved in Proposition 1 of \cite{Bolte}.

\smallskip

In the specific case where the Riemann surface $X$ is defined over a number field,
so that it admits a Belyi map $f: X=H \backslash \H \to 
\widehat\C=\Delta_{p,q,r}\backslash \H$ and a uniformization $X=H \backslash \H$
by a finite index subgroup of a triangle group $\Delta_{p,q,r}$, as in \cite{Cohen},
the automorphic functions approach of \cite{Bolte} to the spectral decomposition
of the Dirac operator can be made more explicit, using the results of \cite{Doran3}
on automorphic forms for triangle groups. 

\smallskip

\begin{lem}\label{SpActDiracSelb}
Let $X$ be a compact Riemann surface of genus $g=g(X)\geq 2$, endowed with the hyperbolic
metric of constant curvature $-1$. Let $\fs$ be a spin structure on $X$ and $D=D_\fs$ the
corresponding Dirac operator, acting on sections of the spinor bundle $\bS=\bS_\fs$ on $X$.
Let $\chi:\tilde\Gamma \to U(1)$ be the character determined by the spin structure $\fs$, as above.
Let $h\in \cC^\infty(\R)$ be an even compactly supported test function with support ${\rm supp}(h) =[-1,1]$
and let $f\in \cS(\R)$ be the Fourier transform $f=\hat h$. The Dirac spectral action then satisfies
\begin{equation}\label{hypSelbDiracSA}
\begin{array}{rl}
\cS_{D,f}(\Lambda)= & \Lambda^2 (g(X)-1) \int_\R r f(r) \coth(\pi r) dr \\[4mm] + & \Lambda
\displaystyle{\sum_{P\in \cC_{\bar\Gamma}} \sum_{\ell=1}^\infty  
\frac{\chi(P^\ell) \, \arccosh(\frac{t_P}{2}) \, h(\Lambda 2\ell \arccosh(\frac{t_P}{2}))}{\sinh(\ell \arccosh(\frac{t_P}{2}))}}
\end{array}
\end{equation}
where $\cC_{\bar\Gamma}$ is the set of $\bar\Gamma$-conjugacy classes.
\end{lem}

\proof According to Theorem 1 of \cite{Bolte}, the Selberg trace formula for the Dirac operator
on a hyperbolic compact Riemann surface is obtained by modifying the case \eqref{Selb2} of the Laplacian
in the following way:
\begin{equation}\label{Selb2Dirac}
\begin{array}{rl}
\displaystyle{\sum_{j=0}^\infty f(\lambda_j)} = & 
\displaystyle{(g(X)-1) \int_\R r f(r) \coth(\pi r)
\, dr} \\[4mm] + & \displaystyle{\sum_{P\in \cC_{\bar\Gamma}} \arccosh(\frac{t_P}{2}) \sum_{\ell=1}^\infty \chi(P^\ell)\, 
\frac{h(2\ell \arccosh(\frac{t_P}{2}))}{\sinh(\ell \arccosh(\frac{t_P}{2}))}} , \end{array}
\end{equation}
where $\{ \lambda_j \} =\Spec(D_\fs)$.
Again, we replace $f(r)$ with $f_\Lambda(r)=f(r/\Lambda)$ and $h(s)$ with
$h_\Lambda(s)=\Lambda\, h(\Lambda s)$, to obtain \eqref{hypSelbDiracSA}.
\endproof

\smallskip
\subsection{Selberg zeta function and the Spectral Action}

An approach to the Selberg trace formula via the Selberg zeta function is described in  \cite{Fisher},
see also \S VII of \cite{Bolte} for the Dirac case. For $\Re(s)>1$ and $\Re(\sigma)>1$, the trace
formula applied to the test function
\begin{equation}\label{testZSelb}
 f(\lambda) = (\lambda^2 + (s-\frac{1}{2})^2)^{-1} - (\lambda^2+(\sigma-\frac{1}{2})^2)^{-1} 
\end{equation} 
gives as the second term in the right-hand-side of \eqref{Selb2Dirac} the expression
$$ \frac{1}{2s-1} \frac{Z_\Gamma^\prime(s)}{Z_\Gamma (s)} - \frac{1}{2\sigma-1} \frac{Z_\Gamma^\prime(\sigma)}{Z_\Gamma(\sigma)}, $$
where $Z_\Gamma(s)$ is the Selberg zeta function
\begin{equation}\label{SelbergZ}
Z_\Gamma (s) = \prod_{P\in \cC_{\bar\Gamma}} \prod_\ell (1-\chi(P) e^{L_P (s+\ell)}),
\end{equation}
where the set $\cC_{\bar\Gamma}$ is identified with the set of primitive closed geodesics on $X=\Gamma\backslash \H$ and$L_P=2\arccosh(\frac{t_P}{2})$ with the geodesic length. 

\smallskip

The Selberg zeta function can sometimes be computed in terms of a coding of 
geodesics via symbolic dynamics and the 
Fredholm determinant of an associated Ruelle transfer operators. For the case
of the modular group $\SL_2(\Z)$ and finite index subgroups, see \cite{ChangMayer}.
Similar results have been obtained for a particular class of Fuchsian triangle groups,
the Hecke triangle groups, see \cite{Mayer}, \cite{Strom}, and generalized to
an algorithm for symbolic coding of geodesics applicable to other Fuchsian groups
in \cite{Pohl}. 

\smallskip

Hecke triangle groups $H_q$ are the Fuchsian triangle groups of the form $\Delta_{2,q,\infty}$. They have a
presentation $H_q=\langle S,T_q\,|\, S^2=(ST_q)^q=1\rangle$, with $Sz=-1/z$ and $T_q z=z+\lambda_q$,
where $\lambda_q=2\cos(\pi/q)$. A compact Riemann surface $X$ that admits a regular tessellation by
hyperbolic $q$-gons also admits a uniformization by a finite index subgroup $H$ of a Hecke group $H_q$, with
a Belyi map given by the projection $f:  H\backslash \H \to \P^1=H_q\backslash \H$, with the single cusp $\{ \infty \}$
of $H_q$ added in the compactification $\P^1$. 

\smallskip

The Selberg zeta function of a Hecke triangle group $H_q=\Delta_{2,q,\infty}$ can be computed explicitly,
in terms of thermodynamic formalism and the appropriate transfer operator, associated to a continued
fraction algorithm, as shown in \cite{Mayer}, \cite{Strom}. More precisely, the Selberg zeta function is obtained,
in the case of the Hecke triangle groups $H_q$ as a ratio of Fredholm determinants
\begin{equation}\label{ZHqFred}
Z_{H_q}(s) = \frac{\det(1-\cL_s)}{\det(1-\cK_s)},
\end{equation}
where $\cL_s$ is the Liouville transfer operator of the continued fraction algorithm of $H_q$, in analogy
to the case of $\SL_2(\Z)$ of \cite{ChangMayer}.
However, the operator $\cL_s$ alone introduces an overcounting, in the case of triangle groups $H_q$, which
is corrected by another transfer operator $\cK_s$. Moreover, in \cite{MoPo} it is shown that the Maass cusp forms
of Hecke triangle groups $H_q$ are solutions of a certain functional equation, generalizing the one of Lewis--Zagier for
the modular group, which characterizes fixed points of the transfer operator associated to the Selberg zeta function.  

\smallskip

A coding of geodesics for the more general triangle groups $\Delta_{p,q,\infty}$ and associated
transfer operators, whose Fredholm determinant is similarly related to the Selberg zeta function,
were obtained in \cite{Fried}. A more general approach to symbolic dynamics and transfer operators, 
for arbitrary $2$-dimensional hyperbolic good orbifolds $X$, is developed in \cite{Pohl}. 

\smallskip

These methods provide an approach to compute the spectral action $\cS_{D,f}(\Lambda)$
with a test function of the form \eqref{testZSelb} using the Selberg zeta function computed via
the transfer operator method.

\smallskip

The transfer operator method in general consists of a construction of a cross section for
the geodesic flow on the hyperbolic surface $X=\Gamma\backslash \H$, where $\Gamma$
is a cocompact Fuchsian group, using a choice of fundamental domains,
so that the first return map determines a discrete dynamical system. This dynamical
system is then encoded on the boundary $\P^1(\R)$ of $\H$ in terms of a family of
finitely many local diffeomorphisms, determined by a coding of geodesics by infinite sequences in 
an alphabet $\Sigma$ in which the first return map becomes the shift map of the symbolic dynamics.
Given the resulting map $F$ describing this boundary dynamics, the associated transfer 
operator (depending on a parameter $\beta\in \C$)
is given by
\begin{equation}\label{transop}
\cL_{F,\beta} f(x) =\sum_{y\in F^{-1}(x)} | F^\prime(y) |^{-\beta} \, f(y).
\end{equation}
We refer the reader to \cite{Pohl} for a detailed construction of the transfer
operator for a class of Fuchsian groups. 

\smallskip
\subsection{Dirac spectral action of Adinkras} 

Given an Adinkra chromotopology $A_{N,k}$ and the associated Riemann surface $X_{N,k}$
with a Belyi map $f: X_{N,k}\to \hat\C$ as in \cite{DILM}, we can consider the spectral triple
$(\cC^\infty(X_{N,k}), L^2(X_{N,k},\bS_\fs), D)$ with $D$ the Dirac operator $D=\Dirac_\fs$
associated to the spin structure $\fs$ determined by the dashing of the Adinkra graph as
in \cite{DILM}. 
By the discussion above and the results of \cite{Cohen}, we know that the Riemann surface
can be uniformized as $X_{N,k} =H \backslash \H$, where $H$ is a finite index subgroup of
the Fuchsian triangle group $\Delta_{N,N,2}$. While in general it is difficult to describe the
subgroup $H$ explicitly, it is still possible to use the method of \cite{Pohl}, applied to the
Fuchsian triangle group $\Delta_{N,N,2}$, as an approach to the computation of the spectral
action, by including the finite coset space $\Delta_{N,N,2}/H$ in the construction of the transfer 
operator, as was done for the finite index subgroups of the modular group in \cite{ChangMayer},
see also \cite{ManMar}, \cite{Mar}. 

\smallskip

More precisely, suppose given a construction as above of a transfer 
operator $\cL_{F,\beta}= \cL_{F,\beta,\Gamma}$ 
based on the coding of the geodesic flow on a hyperbolic surface $X=\Gamma\backslash \H$,
for a given Fuchsian group $\Gamma$. There is a way to obtain from it a transfer
operator for an arbitrary finite index subgroup $H \subset \Gamma$, using the
same boundary dynamics that determines $\cL_{F,\beta,\Gamma}$. This is a simple generalization
of the same construction used in \cite{ChangMayer}, \cite{ManMar}, \cite{Mar} for the case of
$\Gamma=\SL_2(\Z)$. We assume the following general condition: the transfer
operator can be written in terms of local determinations of the function $F$ as
\begin{equation}\label{Llocdetg}
\cL_{F,\beta,\Gamma} f (x) = \sum_E \sum_{s\in \Sigma_E} \chi_{E_s}(x) \, |g_s^\prime(x)|^{-\beta} \, f(g_s x),
\end{equation}
where $F^{-1}(x)=\cup E$ is a union of pairwise disjoint sets $E=\cup_s E_s$, labelled by 
elements $s \in \Sigma_E\subset \Sigma$ in the alphabet of the symbolic coding, 
such that $F|_E(y)=g_s^{-1} y=x$, for elements $g_s\in \Gamma$. 

\begin{lem}\label{Lcoset}
Let $\Gamma$ be a Fuchsian group that has a transfer operator $\cL_{F,\beta,\Gamma}$,
with $F(x)$ the boundary dynamical system providing the coding of geodesics on $\Gamma\backslash \H$,
which satisfies \eqref{Llocdetg}. Let $H\subset \Gamma$ be a finite index subgroup and $X=H\backslash \H$
the corresponding hyperbolic surface. For $P=\Gamma/H$, a coding map for the geodesics on $X$
can be obtained by extending $F(x)$ uniquely to a function $F(x,a)$ with $a\in P$, with transfer
operator of the form
\begin{equation}\label{LP}
\cL_{F,\beta,H\subset \Gamma}f (x,a) =\sum_E \sum_{s\in \Sigma_E} \chi_{E_s}(x) \,
|g_s^\prime(x)|^{-\beta} f(g_sa, g_s a).
\end{equation}
\end{lem}

\proof
Let $P=\Gamma/H$ be the coset space, with the left transitive action of $\Gamma$. We extend
the map $F(x)$ of the boundary dynamics to a map $F(x,a)$, with $a\in P$, by setting
$$ F|_{E_s\times P}(x,a)=(g_s^{-1}x, g_s^{-1}a). $$
Correspondingly we obtain a transfer operator of the form \eqref{LP}.
This transfer operator and the map $F$ considered here provide a coding of 
geodesics on $X=H\backslash \H$ through the identification of this quotient with
the quotient $\Gamma \backslash (\H \times P)$. 
\endproof

\smallskip
\subsection{Supersymmetric Riemann Surfaces}

A Super Riemann Surface $M$ is locally modeled on $\C^{1|1}$, with 
local coordinates $z$ (bosonic) and $theta$ (fermionic). A non-integrable
subbundle $\cD\subset T\C^{1|1}$ is determined by 
$$ D_\theta = \partial_\theta + \theta \partial_z, $$
which satisfies 
$$ [D_\theta, D_\theta]= 2 \partial_z, $$
so that one has $\cD\otimes\cD\simeq TM/\cD$.
We refer the reader to \cite{Ma1}, \cite{Ma2} for a detailed treatment
of the theory of supermanifolds and in particular Super Riemann Surfaces.

\smallskip

It is shown in \cite{DILM} that an {\em odd dashing} on an Adinkra $A_{N,k}$
determines a Super Riemann Surface structure on $X_{N,k}$. 

\smallskip

Thus, one can refine the data of the spectral triple and spectral action discussed above,
based on the Riemann surface $X_{N,k}$, by including also the structure of Super Riemann
Surface. To this purpose, we modify the definition and computation of the spectral action
given above to incorporate the supermanifold structure, by replacing the Selberg trace
formula with a Selberg supertrace formula.

\smallskip
\subsection{The Supersymmetric Selberg trace formula}

A Selberg super trace formula for Super Riemann Surfaces, based on the
Dirac Laplacian, was obtained in \cite{BarMan}. Additional results on the
Selberg super zeta function were obtained in \cite{Gro}, see also \cite{Gro2}.
We consider here the Dirac Laplacian $\Delta = 2 Y D \bar D$ (the case $m=0$
of the family of Dirac Laplacians considered in \cite{BarMan}, \cite{Gro}), where
$-(4Y^2)^{-1}$ is the superdeterminant of the metric tensor on the super upper
half plane $\cS\H$, and $D=D_\theta=\theta \partial_z+\partial_\theta$ and
$\bar D=D_{\bar\theta}=-\partial_{\bar\theta} +\bar\theta \partial_z$. Let
$\{ \lambda^B_j =i r^B_j+1/2 \}$ and $\{ \lambda^F_j=i r^F_j+1/2 \}$ denote, 
respectively, the bosonic and fermionic spectra of $\Delta$. 

\smallskip

Note that the operator $D$ satisfies $D^2 =\partial_z$, so it can be viewed
as a square root of $\partial_z$. This is reflected in the structure of the spectrum,
with respect to the Dirac spectrum on an ordinary Riemann surface. Thus, a
natural type of spectral action functional to consider in this supersymmetric
spectrum is obtained by replacing the ordinary Dirac operator by the
supersymmetric Dirac Laplacian $\Delta$ and the trace by a supertrace.

\smallskip

Let $f$ be a test function with the properties that $f(ix+1/2)$ is in $\cC^\infty(\R)$ with
$f(ix+1/2)\sim O(x^{-2})$ for $x\to \pm \infty$, and with $f(iz+1/2)$ holomorphic for
$|\Im(z)|\leq 1+\epsilon$.

\smallskip

\begin{defn}\label{sSpAct}
For a test function $f$ as above. The supersymmetric spectral action of the Super Riemann
Surface $\cS X =\Gamma \backslash \cS\H$ is given by
\begin{equation}\label{sSA}
\cS_{\cS X, \Delta, f}(\Lambda) =\Tr_s( f(\Delta/\Lambda) ) =\sum_{j=0}^\infty (f(\frac{\lambda_j^B}{\Lambda})-f(\frac{\lambda_j^F}{\Lambda})).
\end{equation}
\end{defn}

\smallskip

One can consider also a slightly different version of the supersymmetric spectral action,
defined by analogy with the Laplacian spectral action discussed above. Let $\tilde f(r)=f(ir+1/2)$
with $f$ a test function as above and define the supersymmetric spectral action as
\begin{equation}\label{tildefSAsusy}
\cS_{\cS X, \Delta, \tilde f}(\Lambda) =\sum_{j=0}^\infty (\tilde f(\frac{r_j^B}{\Lambda})-\tilde f(\frac{r_j^F}{\Lambda})).
\end{equation}
The difference with respect to the previous version lies in rescaling $r^B_j \mapsto \Lambda^{-1} r^B_j$
and $r^F_j \mapsto \Lambda^{-1} r^F_j$ rather than $\lambda^B_j =r_j^B+1/2 \mapsto \Lambda^{-1} \lambda^B_j =\frac{r_j^B+1/2}{\Lambda}$ and similarly $\lambda^F_j \mapsto \Lambda^{-1} \lambda^F_j$. 

\smallskip

Let $h$ denote the Fourier transform
$$ h(t) =\frac{1}{2\pi} \int_\R e^{-itx}\, f(ix+1/2)\, dx, $$
of a test function $f$ chosen as above. Let $G(x,\chi)$ be the function 
\begin{equation}\label{Gxchi}
 G(x,\chi) = h(x)+ h(-x) - (\chi\, e^{-x/2} h(x) + \chi\, e^{x/2} h(-x)). 
\end{equation} 
The supersymmetric spectral action can then be computed in terms
of the Selberg supertrace formula.

\smallskip

\begin{prop}\label{SUSYSpAct}
Let $f$ be a test function as above. The supersymmetric spectral action satisfies
\begin{equation}\label{sSASelb}
\begin{array}{rl}
\cS_{\cS X, \Delta, f}(\Lambda) = &i \Lambda (g(X)-1) \int_\R f(ir+1/2) \tanh(\Lambda \pi r)\, dr \\[4mm]
+ & \displaystyle{ \sum_{\gamma \in \cC(\Gamma)} \sum_{k=1}^\infty \frac{\lambda(\gamma)}{N_\gamma^{1/2}-N_\gamma^{-1/2}} G_\Lambda(\log N_\gamma, \chi(\gamma)), }
\end{array}
\end{equation}
where $G_\Lambda(x,\chi) =h_\Lambda(x)+ h_\Lambda(-x) - (\chi\, e^{-x/2} h_\Lambda(x) 
+ \chi\, e^{x/2} h_\Lambda(-x))$, with the function 
$h_\Lambda(t) = \Lambda e^{-\frac{t}{2} (\Lambda-1)} h(\Lambda t)$.
\end{prop}

\proof
The Selberg supertrace formula is given by (\cite{BarMan}, \cite{Gro}) 
\begin{equation}\label{sTrSelb}
\begin{array}{rl}
\sum_{j=0}^\infty (f(\lambda_j^B)-f(\lambda_j^F)) = & i(g-1) \int_R f(ir+1/2) \tanh(\pi r)\, dr \\[4mm]
+ & \displaystyle{ \sum_{\gamma \in \cC(\Gamma)} \sum_{k=1}^\infty \frac{\lambda(\gamma)}{N_\gamma^{1/2}-N_\gamma^{-1/2}} G(\log N_\gamma, \chi(\gamma)), }
\end{array}
\end{equation}
where the function $G(x,\chi)$ is given by \eqref{Gxchi}.
Here we identify the set $\cC(\Gamma)$ of conjugacy classes of $\Gamma$ with
the oriented primitive closed geodesics and we write 
$\lambda(\gamma)=\ell(\gamma_0)=\log N_{\gamma_0}$ for the length of the unique
element in the class of $\gamma$ such that $\gamma =\gamma_0^m$ for some $m\in \N$.
We write $N_\gamma=\exp(\ell(\gamma))$ for the exponentiated lengths. When scaling the
spectrum of the Dirac Laplacian by $\Delta \mapsto \Delta/\Lambda$, we replace the test
function $f$ with the scaled function $f_\Lambda(\lambda)=f(\lambda/\Lambda)$. Correspondingly,
the Fourier transform $h_\Lambda(t)=(2\pi)^{-1} \int_\R e^{-itx}\, f_\Lambda(ix+1/2)\, dx$ is given by
$h_\Lambda(t) = \Lambda e^{-\frac{t}{2} (\Lambda-1)} h(\Lambda t)$. 
\endproof

The case of the form \eqref{tildefSAsusy} of the supersymmetric spectral action is
similar.

\begin{cor}\label{tildefcase}
Let $f$ be a test function as above. The supersymmetric spectral action in the form  \eqref{tildefSAsusy}
satisfies
\begin{equation}\label{susyS2}
\begin{array}{rl}
\cS_{\cS X, \Delta, \tilde f}(\Lambda) = &i \Lambda (g(X)-1) \int_R \tilde f(r) \tanh(\Lambda \pi r)\, dr \\[4mm]
+ & \Lambda \displaystyle{ \sum_{\gamma \in \cC(\Gamma)} \sum_{k=1}^\infty \frac{\lambda(\gamma)}{N_\gamma^{1/2}-N_\gamma^{-1/2}} G(\Lambda \log N_\gamma, \chi(\gamma)). }
\end{array}
\end{equation}
\end{cor}

\proof The argument is analogous to the previous case, with $\tilde f_\Lambda(r)=\tilde f(\frac{r}{\Lambda})$
and the Fourier transform $h_\Lambda(t) =(2\pi)^{-1}\int_\R e^{-itx} \tilde f_\Lambda(x)\, dx=\Lambda h(\Lambda t)$. 
\endproof

\smallskip
\subsection{Origami curves and the Laplace spectrum}

In addition to the spectral geometry considered above, obtained from the
spectrum of the Dirac operator on the (Super) Riemann surface $X_{N,k}$,
we can also consider another geometry associated to the Adinkra $A_{N,k}$,
namely the origami curve $Y_{N,k}$ considered in Lemma~\ref{Aorigami}.

\smallskip 

In the case of a branched cover $p: Y\to E$ of an elliptic curve $E$, it
is shown in \cite{Mato1}, \cite{Mato2} that it is possible to construct an
infinite set of eigenvalues and eigenfunctions of the Laplacian on $Y$,
which we recall briefly here.

\smallskip

Let $\{ \omega_k \}_{k=1}^g$ be a basis of holomorphic differentials
for $Y$ and let $\Omega$ be the period matrix $\Omega_{jk}=\int_{\beta_j}\omega_k$,
with the normalization $\int_{\alpha_j}\omega_k=\delta_{jk}$ for $\{ \alpha_j,\beta_j\}$
a symplectic basis of $H_1(Y,\Z)$. It is shown in \cite{Mato1}, \cite{Mato2} that each
solution of the equations
\begin{equation}\label{solnsY}
 m_i -\sum_{k=1}^g \Omega_{ik} n_k = N_{ij} (m_j - \sum_{k=1}^g \Omega_{jk} n_k ) 
\end{equation} 
for $(n,m)\in \Z^{2g}$, with $N_{ij} N_{jk}=N_{ik}$ determined by $g-1$ elements, 
determines a metric $g=g^{(n,m)}$ on $Y$. 

\smallskip

One considers then the set of solutions $(n',m')\in \Z^{2g}$ of 
$\omega_{n',m'}=c \omega_{n,m}$, for some $c=c(n,m,n',m')$,
where the $\omega_{n,m}$ are the
primitive differentials 
$$ \omega_{n,m} =\sum_{k=1}^g c_{n,m;k} \, \omega_k, $$
$$ c_{n,m;k} = \pi \sum_{j=1}^g \frac{ m_j - \sum_{\ell =1}^g \bar\Omega_{\ell j} n_\ell }{(\Im(\Omega)^{-1})_{jk}}. $$
As in \cite{Mato2}, we denote by $\cS_{n,m}(\Omega)$ this set of solutions.
This set of solutions in particular includes the $(kn,km)$ for $k\in \Z$, but can 
in general be larger. Each such solution $(n',m')$ determines an eigenvalue
$$ \lambda_{n'm'} = 2 A_{n,m} \left| \frac{m'_i -\sum_{k=1}^g \bar\Omega_{ik} n'_k}{m_i -\sum_{k=1}^g \bar\Omega_{ik} n_k} \right|^2, $$
of the Laplacian $\Delta=\Delta_{g^{(n,m)}}$, where 
$A_{n,m}=\frac{i}{4}\int_Y \omega_{n,m}\wedge \bar\omega_{n,m}$.

\smallskip

Solutions to the equation \eqref{solnsY} can be seen as a set of consistency equations satisfied
by the period matrix of $Y$. It is shown in \cite{Mato2} that a Riemann
surface $Y$ has period matrix satisfying these conditions if and only
if it is a branched cover of an elliptic curve. Thus, in particular, the
construction works for all origami curves. 

\smallskip

Thus, given the origami curve $Y=Y_{N,k}$, associated to an Adinkra graph $A_{N,k}$
as in Lemma~\ref{Aorigami}, one can consider the spectral action functional associated
to the spectrum $\{ \lambda_{n',m'} \}$ constructed in \cite{Mato1}, \cite{Mato2}. Since 
the spectrum considered in the action functional should correspond to a square root
of the Laplacian, we consider the sequence
$$ \rho_{n',m'} = \pm \sqrt{2 A_{n,m}} \, 
\left| \frac{m'_i -\sum_{k=1}^g \bar\Omega_{ik} n'_k}{m_i -\sum_{k=1}^g \bar\Omega_{ik} n_k} \right| $$
for $(n',m')\in \cS_{n,m}(\Omega)$, and we define
\begin{equation}\label{seriesY}
\cS_{(n,m),Y,f}(\Lambda) := \sum_{(n',m')\in \cS_{n,m}(\Omega)} f(\rho_{n',m'}/\Lambda),
\end{equation}
for an even test function $f\in \cS(\R)$. 

\smallskip

These eigenvalues have a structure similar to the spectrum of the Dirac operator
on a torus, hence the computation of the action functional \eqref{seriesY} can be
approached via a Poisson summation formula.

\bigskip

\subsection*{Acknowledgement} The first author is supported by NSF grants 
DMS-1201512 and PHY-1205440. The second author is supported by a
Summer Undergraduate Research Fellowship at Caltech. We thank Kevin Iga 
for a careful reading of the manuscript and for very useful comments and suggestions.

\end{document}